\documentclass[letterpaper,comsoc]{IEEEtran}
\usepackage{cite}
\usepackage{algorithm}
\usepackage{amsmath,amssymb,amsfonts}
\usepackage{algorithmic}
\usepackage{graphicx}
\usepackage{textcomp}
\usepackage{epstopdf}
\usepackage{stfloats}
\usepackage{setspace}
\ifCLASSOPTIONcompsoc
\usepackage[caption=false, font=normalsize, labelfont=sf, textfont=sf]{subfig}
\else
\usepackage[caption=false, font=footnotesize]{subfig}
\usepackage{bm}

\usepackage[T1]{fontenc}
\usepackage{cite}
\usepackage{amsmath,amssymb,amsfonts}
\usepackage{algorithmic}
\usepackage{graphicx}
\usepackage{textcomp}
\usepackage{xcolor}
\usepackage{bm}
\usepackage{changepage}
\usepackage{amsmath}
\usepackage{amstext}
\usepackage{stfloats}
\usepackage{url}

\usepackage{amsmath}

\begin{document}
\markboth{accepted by IEEE TRANSACTIONS ON Wireless COMMUNICATIONS}%
{Shell \MakeLowercase{\textit{et al.}}: Bare Demo of IEEEtran.cls for IEEE Communications Society Journals}
\title{
\huge Channel Estimation for Flexible Intelligent Metasurfaces: From Model-Based Approaches to Neural Operators
	\thanks{
	
	Jian Xiao and Ji Wang are with the Department of Electronics and Information Engineering, College of Physical Science and Technology, Central China Normal University, Wuhan 430079, China (e-mail: jianx@mails.ccnu.edu.cn; jiwang@ccnu.edu.cn). 
	
Qimei Cui is with National Engineering Research Center for Mobile Network Technologies, Beijing University of Posts and Telecommunications, Beijing 100876, China (e-mail: cuiqimei@bupt.edu.cn).

Yucang Yang is with the Department of Network Operation, China Unicom Group, Beijing 100048, China, and also with the Department of Electronic Engineering, Tsinghua University, Beijing 100084, China (e-mail: yangyc620@chinaunicom.cn).
	
Xingwang Li is with the School of Physics and Electronic Information Engineering, Henan Polytechnic University, Jiaozuo 454003, China (e-mail: lixingwang@hpu.edu.cn).

Dusit Niyato is with the College of Computing and Data Science, Nanyang Technological University, Singapore 639798 (e-mails: dniyato@ntu.edu.sg).

Chau Yuen is with the School of Electrical and Electronics Engineering,
Nanyang Technological University, Singapore 639798 (e-mail:
chau.yuen@ntu.edu.sg).

	}
	}
			\author{Jian Xiao, Ji Wang,~\IEEEmembership{Senior Member,~IEEE}, Qimei Cui,~\IEEEmembership{Senior Member,~IEEE}, Yucang Yang, \\Xingwang Li,~\IEEEmembership{Senior Member,~IEEE}, Dusit Niyato, \IEEEmembership{Fellow,~IEEE} and Chau Yuen,~\IEEEmembership{Fellow,~IEEE}}
\maketitle
\begin{abstract}

Flexible intelligent metasurfaces (FIMs) offer a new solution for wireless communications by introducing morphological degrees of freedom, dynamically morphing their three-dimensional shape to ensure multipath signals interfere constructively. However, realizing the desired performance gains in FIM systems is critically dependent on acquiring accurate channel state information across a continuous and high-dimensional deformation space. Therefore, this paper investigates this fundamental channel estimation problem for FIM assisted millimeter-wave communication systems. First, we develop model-based frameworks that structure the problem as either function approximation using interpolation and kernel methods or as a sparse signal recovery problem that leverages the inherent sparsity of millimeter-wave channels. To further improve the estimation capability beyond explicit assumptions in model-based channel estimation frameworks, we propose a deep learning-based framework using a Fourier neural operator (FNO). By parameterizing a global convolution operator in the Fourier domain, we design an efficient FNO architecture to learn the continuous operator that maps FIM shapes to channel responses with mesh-independent properties. Furthermore, we exploit a hierarchical FNO (H-FNO) architecture to efficiently capture the multi-scale features across a hierarchy of spatial resolutions. Numerical results demonstrate that the proposed H-FNO significantly outperforms the model-based benchmarks in estimation accuracy and pilot efficiency, {while also showing superior robustness to hardware imperfections.} In particular, interpretability analysis demonstrates that the proposed H-FNO learns an anisotropic spatial filter adapted to the physical geometry of FIM and decomposes the channel into multi-scale features. 

\end{abstract}

%
\IEEEpeerreviewmaketitle

\section{Introduction}
The relentless pursuit of higher data rates, ubiquitous connectivity, and improved energy efficiency for the sixth-generation (6G) wireless networks has spurred the exploration of novel technologies that manipulate the wireless propagation environment itself \cite{10232975}. Reconfigurable metasurface and other flexible array technologies have garnered significant attention by offering the ability to passively reflect and steer electromagnetic waves, thereby creating favorable channel conditions with reduced hardware complexity \cite{10515204, chen2025remaa}. However, these technologies are predominantly based on rigid planar structures, which limits their ability to adapt to complex propagation environments \cite{10923651, 10659326, 10907789}, especially when deep fading occurs due to destructive multipath interference. Flexible intelligent metasurfaces (FIMs) introduce a revolutionary paradigm by endowing the transceiver with physical adaptability \cite{zhou2024flexible}. An FIM is a conformal array whose radiating elements can be precisely displaced, typically in a direction perpendicular to the surface, allowing the entire structure to morph into a non-planar three-dimensional (3D) shape \cite{li2025flexible}. This additional morphological degree of freedom allows the FIM to actively reconfigure its geometry to ensure that multipath signal components add constructively, thus providing a powerful new mechanism to enhance communication performance beyond what is achievable with rigid arrays \cite{11014597}. 

Recent breakthroughs in material science and microfabrication have led to programmable FIM prototypes capable of real-time and precise surface morphing \cite{li2025flexible}. These advancements have paved the way for optimizing FIM-based communication systems, with initial works focusing on maximizing channel capacity and sensing performance by jointly optimizing the FIM shapes and transmit precoding matrices under the assumption of perfect channel information \cite{10850658, 10922153, 11060929}. However, the practical realization of these gains hinges critically on the ability to acquire the accurate channel. Without a precise channel estimate, the subsequent signal processing, from beamforming to shape optimization, becomes ineffective.

\subsection{Prior Works}  
The existing literature on FIMs can be broadly categorized into two main research thrusts. The first line of research focuses on the hardware design perspective, addressing the fundamental challenges required to physically realize dynamically morphing surfaces, while a second complementary line of research addresses the algorithm perspective, developing the necessary signal processing for various FIM applications. In particular, this paper focuses on the communication design required to effectively enhance wireless systems using FIMs.

\subsubsection{Hardware Design of FIM}  
The progression from rigid to flexible metasurfaces has been driven by significant advancements in material science, fabrication techniques, and dynamic actuation mechanisms. The initial shift from rigid to flexible arrays was pioneered in the optical domain. The authors in \cite{kamali2016decoupling} demonstrated that using conformal dielectric metasurfaces could effectively decouple a electromagnetic function of device from its physical shape. This concept was applied to create ultrathin and conformal devices for applications such as lensing and cloaking \cite{cheng2016all}. 
Concurrently, research focused on developing large-area and low-cost fabrication methods, such as those using nanoslits, to make the production of flexible metasurfaces for multispectral wave manipulation more practical \cite{zhang2019large}. Subsequent research expanded the material platform beyond passive dielectrics to enable active and more functional devices, creating multifunctional interfaces \cite{gal2022flexible, zhou2024flexible}. 

A critical enabler for the FIM discussed in this paper is the ability to dynamically and precisely control the surface shape. {The dynamic morphing capability of FIM is enabled by integrating a grid of micro-actuators into the flexible substrate that houses the radiating elements, in which a central controller sends signals to this actuator grid, allowing for the rapid adjustment of the metasurface into a desired 3D shape. In \cite{ni2022soft}, a liquid metal architecture that relies on the electromagnetic Lorentz force was proposed to implement precise and low-power element displacement of FIM.}
The authors in \cite{bai2022dynamically} demonstrated a dynamically reprogrammable surface capable of self-evolving shape morphing \cite{bai2022dynamically}. More recently, these concepts have been extended into the microwave regime with FIMs that feature shape-guided adaptive programming \cite{li2025flexible}. These collective advancements in materials, fabrication, and particularly in dynamic actuation, provide the physical foundation for the FIM hardware. 

\subsubsection{Communication Design of FIM} 
Complementing the advancements in hardware, a significant body of research has focused on the signal processing required to unlock the potential of reconfigurable antenna arrays. {These technologies follow two main paradigms: electronic tuning of electromagnetic properties, exemplified by the emerging reconfigurable intelligent surfaces (RISs), and physical reconfiguration of antenna locations, e.g., fluid, movable, and pinching-antenna systems \cite{9770295, ning2025movable, Ding2024}. Despite their different mechanisms, both share the critical challenge of channel estimation due to their massive configurable degrees of freedom. Consequently, research has yielded both model-based and data-driven solutions to address this challenge. Model-based approaches exploit physical priors, leveraging specific channel estimation approaches, e.g., sparsity-based compressed sensing and tensor decomposition \cite{10053657, hu2025reconfigurable, 10938032, 10659325, 10709906}. For instance, the authors of \cite{10659325} proposed a tensor decomposition-based effective channel estimation framework for movable antenna systems. In \cite{10709906}, the authors developed a highly sophisticated polar-domain dictionary to accurately reconstruct the near-field channel with wideband beam split effect. Conversely, data-driven methods leveraged neural networks to learn the complex and non-linear channel mappings by employing various deep learning architectures \cite{10533725,10129974, 11018390}.}

Building upon this principle of reconfigurable antennas, FIMs advance this principle by integrating the 3D flexibility with the large aperture and element density characteristic of metasurfaces. At present, research into FIM-aided communication systems is still in its nascent stages. Initial representative works have begun to establish the potential performance gains across different scenarios. In \cite{10850658}, the joint optimization of transmit beamforming and the FIM surface shape has been investigated to proactively manage inter-user interference and enhance multiuser performance for FIM-aided multiuser systems. Furthermore, the authors in \cite{10922153} investigated FIM assisted multi-input multi-output (MIMO) systems, where the joint optimization of the FIM 3D surface shape and the transmit covariance matrix was proposed to maximize channel capacity. In \cite{10910066}, the general models for flexible antenna arrays were established and evaluated, which consider various transformations e.g., bending, folding, and rotation, to characterize the resulting channel variations. Beyond enhancing data rates, the unique capabilities of FIMs have been explored for other applications, most notably wireless sensing. the authors in \cite{11060929} investigated the joint design of FIM surface shapes and transmit waveforms to enhance multi-target sensing performance.

\subsection{Motivations and Contributions}  
A common thread throughout the aforementioned application-focused works is the general assumption that the accurate channel is available. However, acquiring the accurate channel in the continuous and high-dimensional deformation space of FIMs is a fundamental and non-trivial challenge \cite{11014597}. This motivates our work to develop novel and efficient channel estimation techniques specifically tailored for FIM-aided communication systems.
{Specifically, in contrast to the channel estimation in conventional rigid arrays which involves estimating a finite set of channel parameters for a fixed geometry \cite{10053657, hu2025reconfigurable, 10938032, 10659325, 10709906}, the channel in an FIM system is a function of the high-dimensional deformation vector that represents the displacement of all radiating elements \cite{10850658, 10922153}. Hence, FIM channel estimation requires acquiring a continuous and high-dimensional function that maps the infinite space of all possible physical deformations to the corresponding channel response.} The space of possible deformations is continuous and infinite, making it impossible to measure the channel for every possible shape. A naive approach of discretizing this high-dimensional space would lead to an intractable number of pilot transmissions, resulting in prohibitive overhead. Therefore, a core challenge is to develop low-overhead channel estimation schemes that can accurately characterize the channel for the entire continuous deformation space based on a limited number of pilot measurements.

Against the above background, we investigate the channel estimation schemes spanning from classic model-based algorithm design to a novel learning-based framework for FIM systems. Our main contributions are summarized as follows.

\hangafter=1
\setlength{\hangindent}{2em}
$\bullet$ We formulate the channel estimation problem for FIM-aided multiuser millimeter-wave (mmWave) systems, which is based on a sophisticated physical model of FIMs. We specifically address the challenge of accurately characterizing the channel across the entire continuous and high-dimensional space of possible FIM deformations while maintaining low pilot overhead.

\setlength{\hangindent}{2em}
$\bullet$ We begin by developing two distinct model-based frameworks that structure the FIM channel estimation problem in fundamentally different ways. The first is an interpolation-based approach, which includes K-nearest neighbors (KNN) and kernel ridge regression (KRR), framing the problem as non-linear function approximation in the deformation space. The second framework leverages the inherent sparsity of mmWave channels, utilizing orthogonal matching pursuit (OMP) to reconstruct the underlying physical parameters of the channel from a limited set of pilot measurements.

\setlength{\hangindent}{2em}
$\bullet$ We further propose a deep learning-based channel estimation framework to move beyond the explicit model assumptions of conventional approaches, which is built upon a Fourier neural operator (FNO). The proposed FNO can learn the continuous operator that maps between function spaces in the Fourier domain, aligning with the wave-based physics of the channel. To capture multi-scale characteristics of the FIM channel, we further develop a Hierarchical FNO (H-FNO) architecture by integrating a U-shape network architecture, which efficiently extracts both the global large-scale features and the local fine-grained details of the shape-to-channel relationship.

\setlength{\hangindent}{2em}
$\bullet$ Numerical results demonstrate that the proposed H-FNO architecture significantly outperforms the developed model-based benchmarks in terms of both estimation accuracy and pilot efficiency. Furthermore, the generalization performance and zero-shot learning ability of the H-FNO is presented. The interpretability visualization reveals that the H-FNO learns physically-grounded features, including multi-scale channel characteristics and an anisotropic Fourier-domain filter, thereby validating its robustness and effectiveness for practical FIM systems.

\subsection{Organizations and Notations}  

The remainder of this paper is organized as follows. Section II introduces the system model for FIM-aided communications and formulates the channel estimation problem. In Section III, we develop the model-based channel estimation frameworks, covering interpolation and sparsity-based techniques. Furthermore, the proposed learning-based FNO framework is presented in Section IV. Section V provides numerical results of the proposed channel estimation models. Section VI concludes the paper.

\emph{Notations}: Lower-case and upper-case boldface letters denote a vector and a matrix, respectively; $\mathbf{A}^T$, $\mathbf{A}^H$ and $\mathbf{A}^{-1}$ denote the transpose, conjugate transpose and pseudo-inverse of matrix $\mathbf{A}$, respectively; ${\bf{I}}_a$ is a $a \times a$ identity matrix; Symbols $| {\cdot}|$, $\left\| {\cdot} \right\|$, and $\left\| {\cdot} \right\|_{F}$ denote the $\ell_1$, $\ell_2$, and Frobenius norm, respectively; Symbol $\lfloor \cdot \rfloor $ denotes the floor function; Symbol $\odot$ denotes the Hadamard product. Symbol $\propto$ denotes the proportionality relation. Symbol $\langle \cdot, \cdot \rangle$ denotes the inner product between two vectors. Operators $\Re(\mathbf{A})$ and $\Im(\mathbf{A})$ denote the real and imaginary components of the complex-valued $\mathbf{A}$.

\section{System Model and Problem Formulation}
\begin{figure}[t]
	\centerline{\includegraphics[width=3.4in]{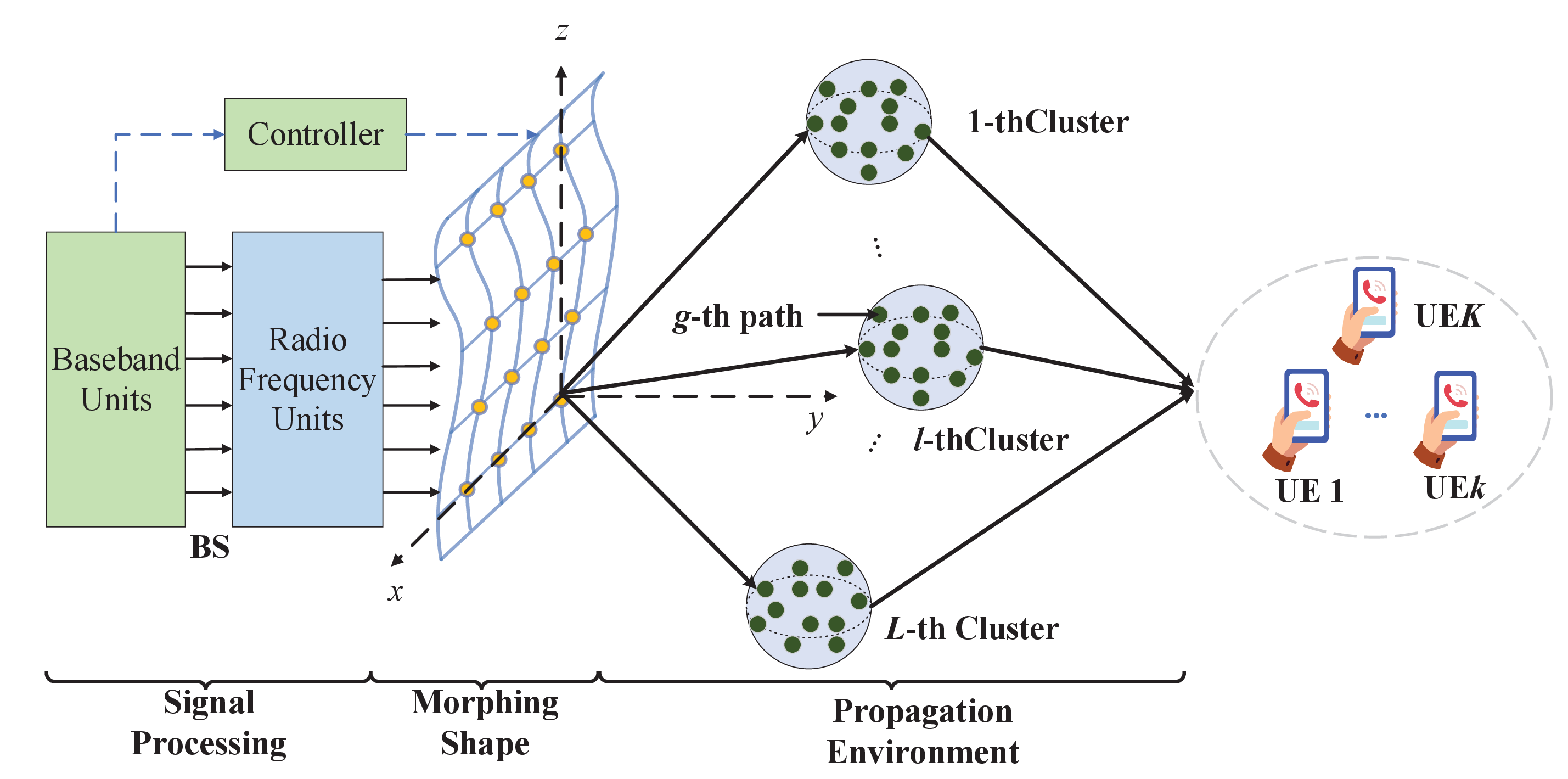}}
	 \caption{FIM assisted multi-user communications.}
	\label{system}
\end{figure}

As shown in Fig.~\ref{system}, we consider downlink communication system where a base station (BS) equipped with an FIM serves $K$ single-antenna user equipment (UE). The FIM consists of $N=N_x \times N_z$ radiating elements arranged as a uniform planar array (UPA) on a flexible substrate. With the aid of controller that can be driven by micro-electromechanical systems (MEMS) or electromagnetic Lorentz force \cite{zhou2024flexible, bai2022dynamically, ni2022soft}, {the FIM can programmatically alter its physical shape to optimize the propagation channels for all users.}

\subsection{FIM Geometry and Deformation} 
To describe an FIM with arbitrary orientation in a global coordinate system, we define a local coordinate system for the FIM. This local system is characterized by three angles: azimuth \(\phi_t \in [0, \pi)\), elevation \(\theta_t \in [0, \pi)\), and spin \(\rho_t \in [0, 2\pi)\). These angles define the local orthonormal basis vectors \(\{\mathbf{i}_t, \mathbf{j}_t, \mathbf{k}_t\}\) of FIM, where \(\mathbf{k}_t\) is the normal vector to the undeformed surface, and \(\{\mathbf{i}_t, \mathbf{j}_t\}\) are orthogonal vectors spanning the surface plane. Following the model in \cite{10922153}, the basis vectors are given by
\begin{equation}
\label{eq:basis_vectors} 
\begin{aligned}
\mathbf{k}_t &= \left[\sin\theta_t\cos\phi_t, \sin\theta_t\sin\phi_t, \cos\theta_t\right]^T, \\
\mathbf{i}_t &= \begin{bmatrix} \sin\phi_t\cos\rho_t - \cos\theta_t\cos\phi_t\sin\rho_t \\ -\cos\phi_t\cos\rho_t - \cos\theta_t\sin\phi_t\sin\rho_t \\ \sin\theta_t\sin\rho_t \end{bmatrix}, \\
\mathbf{j}_t &= \begin{bmatrix} -\sin\phi_t\sin\rho_t - \cos\theta_t\cos\phi_t\cos\rho_t \\ \cos\phi_t\sin\rho_t - \cos\theta_t\sin\phi_t\cos\rho_t \\ \sin\theta_t\cos\rho_t \end{bmatrix}.
\end{aligned}
\end{equation}

The position of the $n$-th element on the undeformed surfac \(\mathbf{p}_n\) is defined relative to the FIM's center, \(\mathbf{p}_c\), by its local two-dimensional (2D) coordinates \((x_n^t, y_n^t)\):
\begin{equation}
\begin{split}\label{vg_condensed}
\mathbf{p}_n = \mathbf{p}_c + x_n^t \mathbf{i}_t + y_n^t \mathbf{j}_t.
\end{split}
\end{equation}

The core feature of the FIM is the controllable deformation. Each element $n$ can be displaced by a distance $\zeta_n \in [-\bar{\zeta}, \bar{\zeta} ]$ along the normal vector \(\mathbf{k}_t\), where $\bar{\zeta}>0$ denotes the maximum range allowed by the unilateral reversible deformation of the FIM \cite{10850658,10922153,11060929}\footnote{{For a physical FIM, the surface shape is described by a continuous function $S(x,z)$, which is governed by the physical properties of the FIM material. The individual displacements are then samples from the function $\zeta_n = S(x_n, z_n)$. These properties naturally enforce inter-element displacement continuity, meaning the displacement of one element is inherently coupled with its neighbors. Furthermore, these properties limit the second spatial derivative of the surface function, thereby imposing a maximum curvature to prevent material fatigue or damage.}}. The collection of these displacements forms the deformation vector {{\(\boldsymbol{\zeta} = [\zeta_1, \dots, \zeta_N]^T \in \mathbb{R}^{N\times 1} \)}}. Final, the post-deformation position of the $n$-th element is given by
\begin{equation}
\begin{split}\label{vg_condensed} 
\tilde{\mathbf{p}}_n = \mathbf{p}_n + \zeta_n \mathbf{k}_t.
\end{split}
\end{equation}

\subsection{Channel Model}
The channel between the $N$-element FIM at the BS and each of the $K$ UEs is distinct and depends on the deformation shape $\boldsymbol{\zeta}$ of FIM. We adopt a clustered multipath channel model, where for each UE $k \in \{1, \dots, K\}$, the channel is formed by $L_k$ scattering clusters, and the $l$-th cluster contributes $G_{k,l}$ individual propagation paths. Following the mmWave channel modeling in 3GPP \cite{3GPP}, the channel $\mathbf{h}_k(\boldsymbol{\zeta}) \in \mathbb{C}^{N\times 1}$ between BS and UE $k$ is given by
\begin{equation}\label{channel}
\mathbf{h}_k(\boldsymbol{\zeta}) = \sum_{l=1}^{L_k} \sum_{g=1}^{G_{k,l}} \alpha_{k,l,g} \, {\mathbf{a}}(\boldsymbol{\zeta}, \mathbf{o}_{k,l,g}),
\end{equation}
where $\mathbf{o}_{k,l,g}$ is the unit-norm direction of departure (AoD) vector for the $g$-th path in the $l$-th cluster to UE $k$. ${\mathbf{a}}(\boldsymbol{\zeta}, \mathbf{o}) \in \mathbb{C}^{N\times 1}$ is the steering vector of the FIM, which is a function of the deformation $\boldsymbol{\zeta}$ and a given AoD $\mathbf{o}$. The AoDs for paths within a single cluster are distributed around a mean cluster angle with a certain angular spread. $\alpha_{k,l,g}$ is the complex gain of the corresponding path, which is  defined as \cite{3GPP}
    \begin{equation}
    \alpha_{k,l,g} = \sqrt{\beta(d_k)} \cdot \frac{1}{\sqrt{\sum_{l'=1}^{L_k} G_{k,l'}}} g_{k,l,g},
    \end{equation}
where $d_k$ is the distance between the BS and UE $k$. $\beta(d_k)= P_0 \left(\frac{d_k}{d_0}\right)^{-\nu}$ is the large-scale path loss coefficient. Here, $P_0$ is the path loss at a reference distance $d_0$, and $\nu$ is the path loss exponent. $g_{k,l,g} \sim \mathcal{CN}(0, 1)$ is the small-scale fading coefficient, modeled as a complex Gaussian random variable.
    
In FIM systems, the steering vector of FIM ${\mathbf{a}}$ can be decomposed into two parts: a rigid component $\mathbf{a}_{\text{rigid}}$ and a flexible component $\mathbf{f}_{\text{flex}}$, i.e., 
\begin{equation}
{\mathbf{a}}(\boldsymbol{\zeta}, \mathbf{o}) = \mathbf{a}_{\text{rigid}}(\mathbf{o}) \odot \mathbf{f}_{\text{flex}}(\boldsymbol{\zeta}, \mathbf{o}).
\end{equation}

The rigid steering vector $\mathbf{a}_{\text{rigid}}$ represents the phase response of the undeformed rigid planar array, which depends only on its fixed orientation and geometry. The $n$-th element of $\mathbf{a}_{\text{rigid}}$ is given by
    \begin{equation}
    \begin{split}\label{vg_condensed} 
      [\mathbf{a}_{\text{rigid}}(\mathbf{o})]_n &= e^{j\kappa \langle \mathbf{p}_n - \mathbf{p}_c, \mathbf{o} \rangle} \\
      &= e^{j\kappa \langle x_n^t \mathbf{i}_t + y_n^t \mathbf{j}_t, \mathbf{o} \rangle},
        \end{split}
    \end{equation}
where $\kappa = 2\pi/\lambda$ denotes the wavenumber and $\lambda$ is the wavelength. 

The flexible phase factor $\mathbf{f}_{\text{flex}}$ captures the additional phase shift introduced solely by the controllable perpendicular deformation, whose $n$-th element can be expressed as
    \begin{equation}
      [\mathbf{f}_{\text{flex}}(\boldsymbol{\zeta}, \mathbf{o})]_n = e^{j\kappa \langle \zeta_n \mathbf{k}_t, \mathbf{o} \rangle}.
    \end{equation}

{In FIM systems, the physical deformation of FIM is typically on the order of a wavelength, e.g., millimeters at mmWave frequencies. Such localized changes are far too small to alter the positions of large-scale scatterers in the environment, e.g., buildings and vehicles. The role of the deformation of FIM is not to change the far-field channel structure but to locally manipulate the phase of the signals at the array by minutely altering the path length to each radiating antenna. This effect is precisely captured in the considered model by the shape-dependent steering vector ${\mathbf{a}}(\boldsymbol{\zeta}, \mathbf{o})$. Hence, the macroscopic scattering environment, which dictates the path gains and angles, is considered independent of the physical deformation of FIM.}

For simplicity and without loss of generality, we assume the undeformed FIM lies on the x-z plane\footnote{{Since any arbitrary orientation of the FIM in a global coordinate system can be described by a rotation matrix. The effect of this rotation on the channel steering vectors is mathematically equivalent to observing the aligned FIM from a correspondingly rotated set of path angles. As the proposed channel estimation frameworks operate on these relative phase shifts, their performance is independent of the initial global orientation, thus ensuring the generality of the proposed algorithms.}}.
{Each radiating antenna of FIM can be flexibly positioned along the surface-perpendicular direction, i.e., the y-axis. This flexibility is consistent with the design of current FIM hardware prototypes \cite{10850658, bai2022dynamically}.}
This simplification corresponds to setting the orientation angles \((\theta_t, \phi_t, \rho_t)\) to \((0, 0, 0)\), which results in the local basis vectors becoming the global Cartesian axes: \(\mathbf{i}_t = [1,0,0]^T\), \(\mathbf{j}_t = [0,0,1]^T\), and the normal vector \(\mathbf{k}_t = [0,1,0]^T\). The undeformed position of the $n$-th element is now \(\mathbf{p}_n = [x_n, 0, z_n]^T\), and the deformation is \(\zeta_n \mathbf{k}_t = [0, \zeta_n, 0]^T\). The final position is thus \(\tilde{\mathbf{p}}_n = [x_n, \zeta_n, z_n]^T\).
Suppose $\theta_{k,l,g}$ and $\phi_{k,l,g}$ are the elevation and azimuth angles of departure for path $(l,g)$ between FIM and the $k$-th user, respectively. The $n$-th element of array response ${\mathbf{a}}$ can be rewritten as
\begin{equation}
\begin{split}\label{vg_condensed}
  \left[\mathbf{a}(\boldsymbol{\zeta}, \theta_{k,l,g}, \phi_{k,l,g})\right]_n = e^{j\frac{2\pi}{\lambda} \left(\tilde{\mathbf{p}}_n \cdot \mathbf{u}_{k,l,g}\right)}, 
  \end{split}
\end{equation}
where $\mathbf{u}_{k,l,g}$ denotes the unit-norm direction vector of path $({l,g})$ and can be expressed as
\begin{equation}
\begin{split}\label{vg_condensed}
\mathbf{u}_{k,l,g} = [\sin\theta_{k,l,g} \cos\phi_{k,l,g}, \sin\theta_{k,l,g} \sin\phi_{k,l,g}, \cos\theta_{k,l,g}]^T.  
\end{split}
\end{equation}

{For a UPA-based FIM with a moderate-scale array aperture, the far-field radiation model is adopted due to the fact that the FIM-users distance is greater than the Rayleigh distance \cite{CM25CKJ}. Considering the ideal mutual coupling and pattern distortion for moderate deformations and element spacing near half a wavelength,} the array response in FIM systems is given by
\begin{equation}
\begin{split}\label{vg_condensed} 
  &\left[\mathbf{a}(\boldsymbol{\zeta}, \theta_{k,l,g}, \phi_{k,l,g})\right]_n = \\
  &e^{j\frac{2\pi}{\lambda} \left(x_n\sin\theta_{k,l,g}\cos\phi_{k,l,g} + \zeta_n\sin\theta_{k,l,g}\sin\phi_{k,l,g} + z_n\cos\theta_{k,l,g}\right)}.   \end{split}
\end{equation}

\subsection{Problem Formulation}
In this work, we investigate a time-division duplexing FIM system, where the downlink channel can be obtained by estimating the uplink channel with the channel reciprocity. Given the deformation shape $\boldsymbol{\zeta}$ of the FIM, the received signal ${\mathbf{y}_{q}}\in {\mathbb{C}^{N\times 1}}$ in the $q$-th slot at the BS is given by
\begin{equation}
\begin{split}\label{signal}
{\mathbf{y}_{q}}&=\sum\limits_{k = 1}^K{\mathbf{{h}}_k(\boldsymbol{\zeta}_q){{{{s}}}_{k,q}}+{\mathbf{w}_{q}}}, 
\end{split}
\end{equation}
where $s_{k,q}$ is transmitted signal at $\text{UE}_k$, and {{${\mathbf{w}_{q}}\sim\mathcal{C}\mathcal{N}(0,{{\sigma}_{w }^{2}\mathbf{I}_N})$}} is complex Gaussian noise.

\begin{figure}[t]
	\centerline{\includegraphics[width=3.2in]{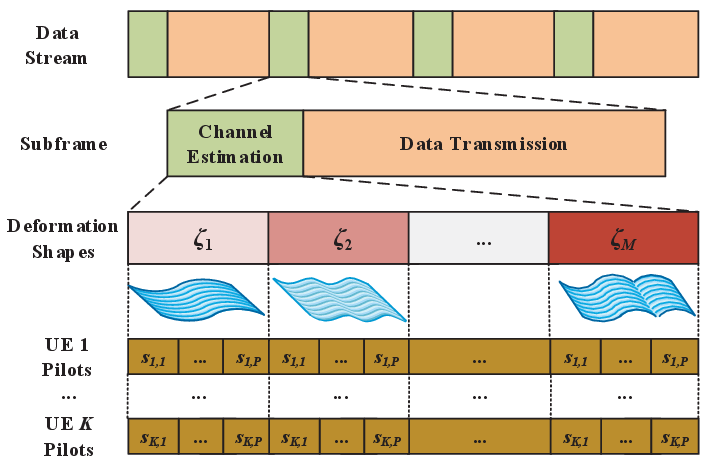}}
	 \caption{Frame protocol for FIM channel estimation.}
	\label{frame}
\end{figure}
As illustrated in Fig.~\ref{frame}, suppose the uplink pilot transmission is divided $M$ subframes, in which each subframe is composed of $P$ pilot slots. {The feasibility of this frame protocol, which involves transmitting pilots from $M$ distinct shapes within a single channel coherence block, is supported by recent hardware prototypes \cite{ni2022soft,bai2022dynamically,li2025flexible}. In particular, the related prototypes of FIM have demonstrated shape-switching times on the order of milliseconds, which are comparable to or faster than the channel coherence time in typical mmWave scenarios.}
In the $m$-th subframe, the collected signal $\mathbf{Y}_m\in {{\mathbb{C}}^{N\times P}}$ at the BS can be expressed as
\begin{equation}\label{sig}
\begin{split}
\mathbf{Y}_m=\sum\limits_{k = 1}^K {{\mathbf{h}_k(\boldsymbol{\zeta}_m)}{\mathbf{s}_k^H}+\mathbf{W}},
\end{split}
\end{equation}
where ${\mathbf{s}}_k = \left[ {{s_{k,1}}, \ldots ,{s_{k,P}}} \right]^T\in {\mathbb{C}^{P\times 1}}$, and $\mathbf{W}={{[{\mathbf{{{w}}}_{1}},\ldots ,{\mathbf{{{w}}}_{P}}]}}\in {{\mathbb{C}}^{N\times P}}$. 		

Considering the widely used orthogonal pilot transmission strategy in multi-user channel estimation and suppose the normalized transmitted power $p_k =1$ at user $k$, we have ${\mathbf{s}}_{{k_1}}^H{{\mathbf{s}}_{{k_2}}} = 0$ and ${\mathbf{s}}_{{k}}^H{{\mathbf{s}}_{{k}}} = P$ for ${k_1} \ne {k_2}(1 \le {k_1},{k_2} \le K)$. In this case, the received signal $\widetilde{\mathbf{{{y}}}}_{k,m}\in {{\mathbb{C}}^{N\times 1}}$ for the $k$-th user can be expressed as
\begin{equation}
\begin{split}
 \widetilde{\mathbf{{{y}}}}_{k,m}= \frac{1 }{P} \mathbf{Y}_m{\mathbf{s}_k}= {\mathbf{h}_k(\boldsymbol{\zeta}_m)}+{\widetilde{\mathbf{w}}_k},
\end{split}
\end{equation}
where $\widetilde{\mathbf{w}}_k=\frac{1 }{ P}\mathbf{W}\mathbf{s}_k$.

Omitting subscript $k$ for notational conciseness, a noisy estimate of the channel at that specific shape is $\hat{\mathbf{h}}(\boldsymbol{\zeta}_m)= \widetilde{\mathbf{{{y}}}}_{m}$. The fundamental problem is to estimate the channel mapping function $\mathbf{h}(\boldsymbol{\zeta})$ for the entire continuous space of deformations $\boldsymbol{\zeta} \in [-\bar{\zeta}, \bar{\zeta}]^N$. A practical approach is to perform a limited number of pilot transmissions, using a set of pre-defined FIM deformation shapes $\{\boldsymbol{\zeta}_1, \boldsymbol{\zeta}_2, \dots, \boldsymbol{\zeta}_{M}\}$.
In this case, the core challenge is to  design a channel estimation procedure that takes the set of noisy measurements $\{\hat{\mathbf{h}}(\boldsymbol{\zeta}_m)\}_{m=1}^{M}$ and produces an accurate estimate $\hat{\mathbf{h}}(\mathbf{\boldsymbol{\zeta}}_{\text{target}})$ for any arbitrary target deformation $\mathbf{\boldsymbol{\zeta}}_{\text{target}}$ with limited pilot overhead $M$.

\section{Model-Based Channel Estimation Approaches for FIM Systems}
In this section, we present and analyze several conventional model-based frameworks for FIM channel estimation. We first explore interpolation-based approaches, which treat the problem as function approximation in the high-dimensional deformation space. We present a practical local linear interpolation method using KNN and a more powerful non-linear approach based on KRR. Subsequently, we shift the paradigm to parametric estimation by leveraging the inherent angular sparsity of the mmWave channel.

\subsection{Interpolation-Based Channel Estimation}
In the channel estimation of FIM systems, the first step is to design a set of $M$ deformation shapes $\{\boldsymbol{\zeta}_m\}_{m=1}^{M}$ in the pilot transmission stage, which is critical for estimation performance. While random shapes are a viable option, a more structured approach is often superior. We propose using a basis set that systematically spans the space of deformations, i.e., a 2D Fourier basis \cite{costa2010unified}, which can be generated using a cosine function and is given by
\begin{equation}
\begin{split}\label{dft} 
\left[\mathbf{y}_{u,v}^{\cos}\right]_n = \bar{\zeta} \cos\left(2\pi\left(\frac{u \cdot i_x(n)}{N_x} + \frac{v \cdot i_z(n)}{N_z}\right)\right),
\end{split}
\end{equation}
where $(i_x(n), i_z(n))$ are the 2D indices of the $n$-th element, and $(u,v)$ are spatial frequency indices. 

By choosing different pairs of $(u,v)$ and using cosine functions, a rich and structured set of orthogonal basis shapes is generated that effectively probe the channel response to different modes of physical deformation. {Compared to random deformation shapes in pilot transmission stage, the superiority of an orthogonal 2D Fourier basis is twofold. First, it guarantees a comprehensive and non-redundant probing of the physical deformation modes of FIMs from low to high spatial frequencies, which is essential for accurately learning the channel function over its continuous domain. Second, this orthogonality systematically minimizes the mutual coherence of the sensing matrix used in the subsequent sparse recovery-based channel estimation algorithms.}

Given the pre-estimated channel set \(\{\hat{\mathbf{h}}(\boldsymbol{\zeta}_m)\}_{m=1}^{M}\) with a given set of \(M\) noisy pilot measurements, a straightforward way to estimate the channel at a new target shape $\mathbf{\boldsymbol{\zeta}}_{\text{target}}$ is through interpolation. The typical interpolation methods can be generally formed as a weighted linear combination of the pilot measurements:
\begin{equation}
\hat{\mathbf{h}}_{\text{interp}}(\boldsymbol{\zeta}_{\text{target}}) = \sum_{m=1}^{M} w_m(\boldsymbol{\zeta}_{\text{target}}, \{\boldsymbol{\zeta}_i\}) \hat{\mathbf{h}}(\boldsymbol{\zeta}_m),
\label{eq:general_interp}
\end{equation}
where $\{w_m\}$ denotes weights of the specific interpolation methods and $\{\boldsymbol{\zeta}_i\}$ represents the entire set of all available pilot deformation shapes. Here, $i$ is a dummy index used to refer to any element within that set.

The core difference between various interpolation techniques lies in the strategy used to compute the weights $\{w_m\}$. The weights $\{w_m\}$ are not static but are a function of the target shape $\boldsymbol{\zeta}_{\text{target}}$ and the set of all pilot shapes $\{\boldsymbol{\zeta}_i\}$. The sophistication of the weighting strategy determines the performance and characteristics of the estimator. 
The simplest scheme is nearest neighbor interpolation, where all weight is concentrated on the single closest pilot shape. The weights are defined using the Kronecker delta function $\delta_{i,j}$, which is equal to 1 if $i=j$ and 0 otherwise. The nearest neighbor interpolation is given by
\begin{equation}
\begin{split}
w_m(\boldsymbol{\zeta}_{\text{target}}, \{\boldsymbol{\zeta}_i\}) = \delta_{m,m^*},
\end{split}
\end{equation}
where $m^* = \arg\min_{i \in \{1,\dots,M\}} \|\boldsymbol{\zeta}_{\text{target}} - \boldsymbol{\zeta}_i\|$ is the index of the nearest pilot shape. This use of the Kronecker delta function results in a piecewise-constant estimate.

By employing a local weighting strategy, linear interpolation operates on a local principle, assuming that the channel function can be locally approximated by a hyperplane. The method first identifies a small neighborhood of pilot points that encloses the target shape $\boldsymbol{\zeta}_{\text{target}}$ and then uses these points to construct a linear estimate.
To build intuition, consider the one-dimensional case where a target shape $\boldsymbol{\zeta}_{\text{target}}$ lies between two pilot shapes $\boldsymbol{\zeta}_a$ and $\boldsymbol{\zeta}_b$ with corresponding channel measurements $\mathbf{h}_a$ and $\mathbf{h}_b$. The linear interpolant for the channel $\mathbf{h}(\boldsymbol{\zeta}_{\text{target}})$ is given by
\begin{equation}
\hat{\mathbf{h}}_{\text{interp}}(\boldsymbol{\zeta}_{\text{target}}) = \mathbf{h}_a + (\mathbf{h}_b - \mathbf{h}_a)\frac{\boldsymbol{\zeta}_{\text{target}} - \boldsymbol{\zeta}_a}{\boldsymbol{\zeta}_b - \boldsymbol{\zeta}_a}.
\end{equation}

In a general $\mathcal{D}$-dimensional space, this concept is extended through barycentric coordinates. First, a simplex, e.g., a triangle in 2D or a tetrahedron in 3D, of $\mathcal{D}+1$ pilot points $\{\boldsymbol{\zeta}_1, \dots, \boldsymbol{\zeta}_{\mathcal{D}+1}\}$ that encloses the target shape $\boldsymbol{\zeta}_{\text{target}}$ is found via Delaunay triangulation \cite{de2008computational}. The target shape can then be uniquely expressed as a linear combination of the simplex vertices:
\begin{align}\label{FFT} 
&\boldsymbol{\zeta}_{\text{target}} = \sum_{m=1}^{\mathcal{D}+1} \lambda_m \boldsymbol{\zeta}_m \\
& \text{s.t.} \quad \sum_{m=1}^{\mathcal{D}+1} \lambda_m = 1 \text{ and } \lambda_m \ge 0, \nonumber
\end{align}
where the coefficients $\{\lambda_m\}$ are the barycentric coordinates and serve as the weights in the general interpolation framework. If $\boldsymbol{\zeta}_m$ is a vertex of the enclosing simplex, $w_m(\boldsymbol{\zeta}_{\text{target}}) = \lambda_m$, otherwise $w_m(\boldsymbol{\zeta}_{\text{target}}) = 0$.
The interpolated channel is then simply $\hat{\mathbf{h}}_{\text{interp}}(\boldsymbol{\zeta}_{\text{target}}) = \sum_{m=1}^{\mathcal{D}+1} \lambda_m \hat{\mathbf{h}}(\boldsymbol{\zeta}_m)$. However, this approach faces significant practical challenges in the high-dimensional space of FIM deformations. For an FIM with $N$ antennas, the deformation space is $N$-dimensional. Constructing a single simplex requires at least $N+1$ pilot shapes. This requirement, $M \ge N+1$, often makes rigorous linear or cubic interpolation computationally infeasible for FIMs where the number of pilot measurements $M$ is typically much smaller than the number of antennas $N$.
The infeasibility of performing a rigorous Delaunay triangulation in the high-dimensional FIM deformation space motivates a more practical and local approximation of linear interpolation. Instead of constructing a global triangulation, we propose a method based on finding a local linear representation of the target shape using its KNN from the set of pilot shapes \cite{hastie2009elements}. 
The KNN-based local linear approach elegantly bypasses the curse of dimensionality associated with global triangulation. By operating in a small, local neighborhood, it remains computationally efficient and numerically stable, while preserving the core principle of linear combination that underlies interpolation methods. 

\subsection{Kernel Ridge Regression-Based Channel Estimation}
While KNN provides a simple heuristic for interpolation, it lacks a strong theoretical foundation for optimality. Kernel-based methods offer a powerful and principled framework for non-linear function approximation, rooted in the theory of reproducing kernel Hilbert spaces (RKHS) \cite{bishop2006pattern}. A prominent and practical example of this approach is kernel ridge regression (KRR) \cite{hastie2009elements}. The KRR models the unknown channel function $\mathbf{h}(\boldsymbol{\zeta})$ as an element of the RKHS, where the estimation for any target shape $\mathbf{\boldsymbol{\zeta}}_{\text{target}}$ is represented as a linear combination of kernel functions evaluated at the pilot locations. The estimate is given by
\begin{equation}
\hat{\mathbf{h}}_{\text{KRR}}(\mathbf{\boldsymbol{\zeta}}_{\text{target}}) = \sum_{m=1}^{M} \mathbf{c}_m k(\mathbf{\boldsymbol{\zeta}}_{\text{target}}, \boldsymbol{\zeta}_m),
\end{equation}
where $k(\cdot, \cdot)$ is a symmetric and positive-definite kernel function that measures the similarity between two shape vectors, and $\mathbf{c}_m \in \mathbb{C}^{N}$ is the coefficient vector to be determined. {The set of pilot shapes $\boldsymbol{\zeta}_m$ is also generated from the 2D Fourier basis, in which the systematic and non-redundant sampling of the deformation space provided by the Fourier basis helps to form a well-conditioned kernel matrix.}

A common and effective choice for the kernel is the Gaussian or radial basis function (RBF) kernel \cite{hastie2009elements}, which can be expressed as
\begin{equation}
k(\boldsymbol{\zeta}_a, \boldsymbol{\zeta}_b) = e^{\left(-\gamma \|\boldsymbol{\zeta}_a - \boldsymbol{\zeta}_b\|^2\right)},
\end{equation}
where $a$ and $b$ denote different FIM deformation shapes. The hyperparameter $\gamma > 0$ controls the bandwidth or length-scale of kernel. {Specifically, $\gamma$ encodes a prior assumption about the smoothness of the channel function. A small $\gamma$ implies a long length-scale, assuming the channel varies smoothly with deformation, while a large $\gamma$ implies a short length-scale, allowing the function to change more rapidly.}

The unknown coefficient matrix $\mathbf{C} = [\mathbf{c}_1, \dots, \mathbf{c}_{M}]^T \in \mathbb{C}^{M \times N}$ is found by solving a regularized least-squares problem. The objective is to find coefficients that minimize the estimation error on the pilot measurements, while also keeping the norm of the coefficients small to prevent overfitting. The Tikhonov-regularized objective function is given by
\begin{equation}
\min_{\mathbf{C}} \quad \|\mathbf{H}_{\text{pilots}} - \mathbf{K} \mathbf{C}\|_F^2 + \lambda \|\mathbf{C}\|_F^2,
\end{equation}
where $\mathbf{H}_{\text{pilots}} = [\hat{\mathbf{h}}(\boldsymbol{\zeta}_1), \dots, \hat{\mathbf{h}}(\boldsymbol{\zeta}_{M})]^T \in \mathbb{C}^{ M \times N}$ is the matrix of all pilot channel measurements. $\mathbf{K} \in \mathbb{R}^{M \times M}$ is the kernel matrix with entries $[\mathbf{K}]_{ij} = k(\boldsymbol{\zeta}_i, \boldsymbol{\zeta}_j)$. {The regularization parameter $\lambda \ge 0$ controls the trade-off between data fidelity and the smoothness prior set. A larger $\lambda$ imposes a stronger penalty on model complexity, enforcing a smoother solution that prevents overfitting to noisy pilot measurements. The optimal value of $\lambda$ cannot be determined analytically. In this work, it is selected using standard model selection techniques by cross-validation.}

This optimization problem has a well-known closed-form solution for the coefficients, which can be expressed as \cite{hastie2009elements}
\begin{equation}
\mathbf{C} = \left(\mathbf{K} + \lambda \mathbf{I}_M\right)^{-1} \mathbf{H}_{\text{pilots}}.
\end{equation}

Finally, to estimate the channel at a new target shape $\mathbf{\boldsymbol{\zeta}}_{\text{target}}$, we first compute the $M \times 1$ vector of kernel similarities, $\mathbf{k}_{\text{target}}$, where $[\mathbf{k}_{\text{target}}]_m = k(\mathbf{\boldsymbol{\zeta}}_{\text{target}}, \boldsymbol{\zeta}_m)$. The final channel estimate is then synthesized as
\begin{equation}
\begin{split}\label{vg_condensed} 
\hat{\mathbf{h}}_{\text{KRR}}(\mathbf{\boldsymbol{\zeta}}_{\text{target}}) &= \mathbf{k}_{\text{target}}^T \mathbf{C} \\
&= \mathbf{k}_{\text{target}}^T \left(\mathbf{K} + \lambda \mathbf{I}\right)^{-1} \mathbf{H}_{\text{pilots}}.
\end{split}
\end{equation}

Compared to KNN, KRR finds an optimal set of weights by solving a global optimization problem. KRR provides a numerically stable and computationally efficient method, requiring only the inversion of a small and well-behaved $M \times M$ kernel matrix. However, the interpolation-based channel estimation method does not leverage prior information about the channel.

\subsection{Sparsity-Based Parametric Channel Estimation}
The parametric channel estimation framework shifts the paradigm from interpolating channel vectors to reconstructing the underlying physical parameters, which is particularly effective when the number of propagation paths is small.
The channel model in \eqref{channel} represents the channel vector $\mathbf{h}(\mathbf{\boldsymbol{\zeta}})$ as a linear combination of $L \times G$ steering vectors. In many practical scenarios, especially at higher frequencies, the number of significant multipath clusters is small \cite{7914742}. This implies that the channel is sparse in the angular domain. We can create an overcomplete dictionary $\mathbf{A}(\mathbf{\boldsymbol{\zeta}}) \in \mathbb{C}^{N \times D}$ whose columns are steering vectors corresponding to a fine grid of $D$ possible angles of arrival, where $D \gg L \times G$. The channel can then be written as
\begin{equation}
\begin{split}\label{vg_condensed} 
\mathbf{h}(\mathbf{\boldsymbol{\zeta}}) = \mathbf{A}(\mathbf{\boldsymbol{\zeta}}) \mathbf{x}, 
\end{split}
\end{equation}
where $\mathbf{x} \in \mathbb{C}^D$ is a sparse vector with only $L \times G$ non-zero entries, representing the complex gains of the active paths.

The goal is to recover the sparse vector $\mathbf{x}$ from the low-dimensional measurements $\{\hat{\mathbf{h}}(\boldsymbol{\zeta}_m)\}_{m=1}^{M}$ with the measurement noise $\mathbf{n} \sim \mathcal{CN}(0, \sigma^2 \mathbf{I})$. We can stack the measurements into a single system of equations:
\begin{equation}
\begin{split}\label{vg_condensed} 
\hat{\mathbf{h}}_{\text{stack}}=\begin{bmatrix} \hat{\mathbf{h}}(\mathbf{\boldsymbol{\zeta}}_1) \\ \vdots \\ \hat{\mathbf{h}}(\mathbf{\boldsymbol{\zeta}}_{M}) \end{bmatrix} = \begin{bmatrix} \mathbf{A}(\mathbf{\boldsymbol{\zeta}}_1) \\ \vdots \\ \mathbf{A}(\mathbf{\boldsymbol{\zeta}}_{M}) \end{bmatrix} \mathbf{x} + \mathbf{n} = \mathbf{\Phi} \mathbf{x} + \mathbf{n}. 
\end{split}
\end{equation}

This is a standard compressed sensing problem, which can be solved using various sparse recovery algorithms. Orthogonal matching pursuit (OMP) is an efficient greedy algorithm for this sparse channel estimation task \cite{tropp2007signal}, which iteratively identifies the support set of the sparse vector $\mathbf{x}$, i.e., the indices of the non-zero entries. 
{In \textbf{Algorithm \ref{alg:omp_channel_est}}, the OMP-based sparse channel reconstruction framework is presented for FIM systems.} Let $\hat{\mathbf{x}}$ be the recovered sparse parameter vector, the channel for any continuous shape $\mathbf{\boldsymbol{\zeta}}_{\text{target}}$ can be synthesized as
$\hat{\mathbf{h}}_{\text{OMP}}(\mathbf{\boldsymbol{\zeta}}_{\text{target}}) = \mathbf{A}(\mathbf{\boldsymbol{\zeta}}_{\text{target}}) \hat{\mathbf{x}}$. This approach elegantly solves both the low-overhead and continuous-space estimation challenges by moving from direct channel interpolation to physical model reconstruction.
{{\begin{algorithm}[!t]
\caption{  OMP-Based Parametric Channel Estimation}
\begin{algorithmic}[1]
 \REQUIRE 
  Observed channel vector $\hat{\mathbf{h}}_{\text{stack}}$, Dictionary matrix $\mathbf{\Phi}$, \\
 \quad     $\text{ }$ Sparsity level $S = L \times G$.

\ENSURE 
$\hat{\mathbf{x}}$: Estimated sparse vector

\STATE Initialize residual: $\mathbf{r}_0 \gets \hat{\mathbf{h}}_{\text{stack}}$
\STATE Initialize support set: $\mathcal{S}_0 \gets \emptyset$

\FOR{$s = 1$ to $S$}
    \STATE Compute correlation: $c_j \gets |\mathbf{\Phi}_j^H \mathbf{r}_{s-1}|, \forall j$
    \STATE Select index: $j_s \gets \arg\max_j c_j$
    \STATE Update support: $\mathcal{S}_s \gets \mathcal{S}_{s-1} \cup \{j_s\}$
    \STATE Solve least squares: $\mathbf{x}_{\mathcal{S}_s} \gets (\mathbf{\Phi}_{\mathcal{S}_s})^{-1} \hat{\mathbf{h}}_{\text{stack}}$
    \STATE Update residual: $\mathbf{r}_s \gets \hat{\mathbf{h}}_{\text{stack}} - \mathbf{\Phi}_{\mathcal{S}_s} \mathbf{x}_{\mathcal{S}_s}$
\ENDFOR

\STATE Construct final solution: 
$\hat{\mathbf{x}} \gets 
\begin{cases} 
\mathbf{x}_{\mathcal{S}_S}[i], & \text{if } j \in \mathcal{S}_S, \\
0, & \text{otherwise}.
\end{cases}$

\RETURN $\hat{\mathbf{x}}$
\end{algorithmic}
\label{alg:omp_channel_est}
\end{algorithm}}}

The performance of OMP is theoretically guaranteed by the properties of the sensing dictionary \(\mathbf{\Phi} \in \mathbb{C}^{MN \times D}\), which is constructed from the pilot shapes \(\{\boldsymbol{\zeta}_m\}\) and a grid of potential angles. A crucial metric is the mutual coherence \(\mu(\mathbf{\Phi})\), which measures the maximum similarity between any two distinct atoms in the dictionary:
\begin{equation}
\begin{split}\label{vg_condensed} 
\mu(\mathbf{\Phi}) = \max_{i \neq j} \frac{|\boldsymbol{\phi}_i^H \boldsymbol{\phi}_j|}{\|\boldsymbol{\phi}_i\|_2 \|\boldsymbol{\phi}_j\|_2}. 
\end{split}
\end{equation}

A low mutual coherence is a sufficient condition for OMP to guarantee exact recovery of the sparse signal in the noiseless case, and it ensures stability and robustness in the presence of noise. Since the sensing dictionary \(\mathbf{\Phi}\) is constructed from the pilot shapes \(\{\boldsymbol{\zeta}_m\}\), we can optimize these shapes to minimize the mutual coherence, thereby creating the best possible conditions for OMP. The optimization problem is given by
\begin{align}
&\{ \boldsymbol{\zeta}_m^* \}_{m=1}^{M} = \arg\min_{\{ \boldsymbol{\zeta}_m \}} \mu \bigl( \boldsymbol{\Phi}(\{ \boldsymbol{\zeta}_m \}) \bigr)  \\
&\quad \text{s.t.} \ -\zeta \le [\boldsymbol{\zeta}_m]_n \le \zeta, \ \forall m, n, \nonumber
\end{align}
{where $\bar{\zeta}$ represents the maximum possible displacement that the hardware actuators can provide without entering a non-linear and unpredictable response regime.} 

To find a high-quality solution to the coherence minimization problem without exhaustive search, a constructive greedy algorithm can be designed that builds the set of optimal pilot shapes sequentially. At each step, it selects a new shape from a large pool of candidates that, when added to the already selected set, results in the minimal possible mutual coherence for the newly formed dictionary. {This approach ensures that each added pilot shape is maximally complementary to the existing ones in terms of distinguishing the dictionary atoms. The detailed procedure is presented in \textbf{Algorithm \ref{alg:2}}.}
\begin{algorithm}[!t]
\caption{  Coherence-Minimal Pilot Selection Optimization}
\begin{algorithmic}[1]

 \REQUIRE 
  Number of pilots \(M\), Candidate shape pool \(\mathcal{Y}_{\text{cand}}\).
\ENSURE 
Optimized pilot shape set \(\mathcal{Y}_{\text{opt}}\).

\STATE Initialize \(\mathcal{Y}_{\text{opt}} \leftarrow \emptyset\).
\STATE Select an initial shape \(\boldsymbol{\zeta}_1\) from \(\mathcal{Y}_{\text{cand}}\) and set \(\mathcal{Y}_{\text{opt}} \leftarrow \{\boldsymbol{\zeta}_1\}\).
\FOR{\(m=2\) to \(M\)}
\STATE Find the next best shape \(\boldsymbol{\zeta}_{\text{next}}\): \\
 \quad \(\boldsymbol{\zeta}_{\text{next}} = \arg \min_{\boldsymbol{\zeta} \in \mathcal{Y}_{\text{cand}} \setminus \mathcal{Y}_{\text{opt}}} \mu(\mathbf{\Phi}(\mathcal{Y}_{\text{opt}} \cup \{\boldsymbol{\zeta}\}))\)\STATE Update \(\mathcal{Y}_{\text{opt}} \leftarrow \mathcal{Y}_{\text{opt}} \cup \{\boldsymbol{\zeta}_{\text{next}}\}\).
\ENDFOR
\STATE \textbf{return} \(\mathcal{Y}_{\text{opt}}\)
\end{algorithmic}
\label{alg:2}
\end{algorithm}

\textbf{Remark 1:} 
While an iterative greedy search can find a pilot set optimized for minimal coherence, it is often computationally prohibitive. {Therefore, we develop a computationally efficient yet highly effective heuristic using the 2D Fourier basis from Eq. \eqref{dft} to generate the pilot shapes. The effectiveness of this approach is explained by the inner product that defines mutual coherence \(\boldsymbol{\phi}_i^H \boldsymbol{\phi}_j = \sum_{m=1}^{M} \mathbf{a}(\boldsymbol{\zeta}_m, \theta_i, \phi_i)^H \mathbf{a}(\boldsymbol{\zeta}_m, \theta_j, \phi_j)\). This summation serves as a discrete approximation of an integral over the deformation space. When the pilot shapes  \(\{\boldsymbol{\zeta}_m\}\) are chosen from an orthogonal basis like the Fourier basis, they act as orthogonal weighting functions in this sum. According to Fourier analysis \cite{costa2010unified}, this process is analogous to projecting one function onto another using an orthogonal basis, which drives the resulting inner product towards zero for non-identical functions.} This systematically decorrelates the dictionary atoms and directly minimizes the mutual coherence.

{While the OMP framework provides a computationally efficient approach to sparse recovery, its performance is sensitive to noise and critically depends on the true channel paths aligning with the predefined angular grid, a problem known as basis mismatch. To establish a more robust sparse baseline, we frame the channel estimation problem within a sparse Bayesian learning (SBL) framework. SBL overcomes the limitation of OMP by treating the sparse channel vector $\mathbf{x}$ as a random variable and learning its statistical properties directly from the measurement data \cite{tipping2001sparse}. The SBL framework recalls the same linear model as in Eq. (28), i.e., $\hat{\mathbf{h}}_{\text{stack}} = \mathbf{\Phi}\mathbf{x} + \mathbf{n} \label{eq:sbl_model}$, we define the likelihood function as
\begin{align}
p(\hat{\mathbf{h}}_{\text{stack}} \mid \mathbf{x}; \sigma^2) = (\pi\sigma^2)^{-MN} e^{\left(-\frac{1}{\sigma^2} \|\hat{\mathbf{h}}_{\text{stack}} - \mathbf{\Phi}\mathbf{x}\|_2^2\right)}. \label{eq:sbl_likelihood}
\end{align}}

{To promote sparsity, SBL places a hierarchical prior on $\mathbf{x}$. Specifically, each element $x_d$ of $\mathbf{x}$ is assumed to be drawn from a zero-mean Gaussian distribution, with its own distinct variance $\alpha_d$, which can be expressed as
    \begin{equation}
    \begin{split} p(\mathbf{x} \mid \boldsymbol{\alpha}) &= \prod_{d=1}^{D} \mathcal{CN}(x_d \mid 0, \alpha_d) \\
&= (\pi)^{-D} \left(\prod_{d=1}^{D} \alpha_d\right)^{-1} e^{\left(-\mathbf{x}^H \mathcal{A}^{-1} \mathbf{x}\right)}, \label{eq:sbl_prior}
    \end{split}
    \end{equation}
where $\boldsymbol{\alpha} = [\alpha_1, \ldots, \alpha_D]^T$ is a vector of hyperparameters, and $\mathcal{A} = \text{diag}(\boldsymbol{\alpha})$. The sparsity is achieved because if a hyperparameter $\alpha_d$ is driven to zero during the learning process, the posterior probability of the corresponding coefficient $x_d$ will be sharply peaked at zero.}

{The SBL algorithm iteratively estimates the posterior distribution of $\mathbf{x}$ while simultaneously learning the hyperparameters $\boldsymbol{\alpha}$ and the noise variance $\sigma^2$ by maximizing the marginal likelihood $p(\hat{\mathbf{h}}_{\text{stack}} \mid \boldsymbol{\alpha}, \sigma^2)$. A common approach to solve this is using an expectation-maximization algorithm. The posterior distribution of $\mathbf{x}$ given the hyperparameters is Gaussian, and the corresponding mean $\boldsymbol{\mu}_x$ and covariance $\boldsymbol{\Sigma}_x$ of $\mathbf{x}$ can be expressed as
\begin{align}
\boldsymbol{\mu}_x = \frac{1}{\sigma^2}\boldsymbol{\Sigma}_x\mathbf{\Phi}^H\hat{\mathbf{h}}_{\text{stack}}. \label{eq:sbl_posterior_mean}
\end{align}
\begin{align}
\boldsymbol{\Sigma}_x = \left(\frac{1}{\sigma^2}\mathbf{\Phi}^H\mathbf{\Phi} + \mathcal{A} ^{-1}\right)^{-1}. \label{eq:sbl_posterior_cov}
\end{align}}

{The posterior mean $\boldsymbol{\mu}_x$ serves as the estimate for the sparse vector $\mathbf{x}$. In each iteration, after computing the posterior statistics, the hyperparameters are updated using the following rules \cite{tipping2001sparse}:
\begin{equation}
\alpha_d^{\text{new}} = \left|(\boldsymbol{\mu}_x)_d\right|^2 + (\boldsymbol{\Sigma}_x)_{d,d} \quad \text{for } d=1,\ldots,D, \label{eq:sbl_alpha_update}
\end{equation}
\begin{equation}
(\sigma^2)^{\text{new}} = \frac{\|\hat{\mathbf{h}}_{\text{stack}} - \mathbf{\Phi}\boldsymbol{\mu}_x\|_2^2 + \text{Re}\left\{\text{Tr}\left(\mathbf{\Phi}\boldsymbol{\Sigma}_x\mathbf{\Phi}^H\right)\right\}}{MN}. \label{eq:sbl_sigma_update}
\end{equation}}

{Upon convergence, many of the values in $\boldsymbol{\alpha}$ will approach zero. We take the final posterior mean $\hat{\mathbf{x}} = \boldsymbol{\mu}_x$ as the recovered sparse vector. The channel for any continuous target shape $\zeta_{\text{target}}$ can then be synthesized in the same way as OMP, i.e., $\hat{\mathbf{h}}_{\text{SBL}}(\mathbf{\boldsymbol{\zeta}}_{\text{target}}) = \mathbf{A}(\mathbf{\boldsymbol{\zeta}}_{\text{target}})\hat{\mathbf{x}}$. By learning the priors from the data, SBL provides a more principled and robust solution to the sparse recovery problem, making it a strong baseline for comparison.}

The above conventional model-based channel estimation frameworks, while foundational, exhibit certain inherent limitations when confronted with the high complexity of the FIM channel estimation problem. The interpolation-based approach, provides the stable linear estimate, yet the fundamental relationship between the FIM deformation vector $\boldsymbol{\zeta}$ and the resulting channel vector $\mathbf{h}(\boldsymbol{\zeta})$ is highly non-linear. Consequently, the linear estimator or kernel-based models are inherently suboptimal and cannot fully capture the intricate channel dynamics. On the other hand, sparsity-based parametric methods based on sparse recovery are constrained by rigid assumptions that may not hold in practice. Their efficacy critically hinges on two conditions: 1) the channel must be sufficiently sparse, which is often not the case in rich scattering environments; and {2) the true path angles must align with a predefined discrete grid $\mathbf{A}(\boldsymbol{\zeta})$ in parametric methods, making the methods highly susceptible to basis mismatch errors. In particular, the true number of paths $S=L\times G$ is unknown to the estimators in practical communication systems, while the performance of parametric methods is critically dependent on this value.} These limitations motivate the development of a more flexible and data-driven approach.

\section{Learning-Based {Channel} Estimation with Fourier Neural Operator for FIM Systems}
To overcome the aforementioned limitations of model-based channel estimation approaches, in this section, we develop a learning-based framework that moves beyond explicit model assumptions of linearity or sparsity. Specifically, we first provide a theoretical justification for learning-based channel estimation from a Bayesian inference perspective. Then, we reframe the channel estimation problem as one of learning a continuous and non-linear operator that maps any given FIM deformation shape to its corresponding channel response. For this purpose, we propose an H-FNO architecture uniquely suited for this task due to its inherent mesh-independence and its design for learning solutions to physical systems governed by partial differential equations (PDEs). {In fact, the propagation of electromagnetic waves is governed by Maxwell's equations, a system of PDEs. The resulting wireless channel can thus be viewed as the solution to the underlying wave equation, where the geometric shape of the FIM, parameterized by the deformation $\mathbf{\boldsymbol{\zeta}}$, acts as a critical boundary condition. Consequently, the task of estimating the channel map ${\mathbf{h}}(\mathbf{\boldsymbol{\zeta}})$ is equivalent to learning the solution operator of this governing PDE. The FNO architecture is specifically designed to approximate such operators that map between infinite-dimensional function spaces, making it a natural and powerful framework for this physics-based problem.}

\subsection{FIM Channel Estimation from a Bayesian Perspective}

In essence, channel estimation can be framed as a Bayesian problem of inferring the most plausible channel function $\bar{\mathbf{h}}$, given a set of noisy measurements, $\{\hat{\mathbf{h}}(\boldsymbol{\zeta}_m)\}_{m=1}^{M}$. According to Bayes' theorem, the posterior probability distribution of the channel function is given by \cite{poor2013introduction}
\begin{equation}
p\left(\bar{\mathbf{h}} \mid \{\hat{\mathbf{h}}(\boldsymbol{\zeta}_m)\}\right) \propto p\left(\{\hat{\mathbf{h}}(\boldsymbol{\zeta}_m)\} \mid \bar{\mathbf{h}}\right) p(\bar{\mathbf{h}}),
\end{equation}
where the likelihood term $p(\{\hat{\mathbf{h}}(\boldsymbol{\zeta}_m)\} \mid \bar{\mathbf{h}})$ is determined by the noise model. 

The goal of channel estimation is to find $\bar{\mathbf{h}}$, which corresponds to the Maximum A Posteriori (MAP) estimate \cite{poor2013introduction}. The MAP estimator maximizes the posterior probability, which is equivalent to minimizing the negative log-posterior:
\begin{equation}
\bar{\mathbf{h}}_{\text{MAP}} = \arg\min_{\bar{\mathbf{h}}} \left( -\log p\left(\{\hat{\mathbf{h}}(\boldsymbol{\zeta}_m)\} \mid \bar{\mathbf{h}}\right) - \log p(\bar{\mathbf{h}}) \right).
\end{equation}

{Given the AWGN model in Eq. \eqref{signal}, the likelihood function is given by
\begin{equation}
p(\{\hat{\mathbf{h}}(\boldsymbol{\zeta}_m)\} \mid \bar{\mathbf{h}}) = \prod_{m=1}^{M} \frac{1}{(\pi \sigma_w^2)^N} e^{-\frac{\|\hat{\mathbf{h}}(\boldsymbol{\zeta}_m) - \bar{\mathbf{h}}(\boldsymbol{\zeta}_m)\|_2^2}{\sigma_w^2}}.
\end{equation}
Taking the negative logarithm of the likelihood gives
\begin{equation}
-\log p(\{\hat{\mathbf{h}}(\boldsymbol{\zeta}_m)\} \mid \bar{\mathbf{h}}) = \sum_{m=1}^{M} \frac{\|\hat{\mathbf{h}}(\boldsymbol{\zeta}_m) - \bar{\mathbf{h}}(\boldsymbol{\zeta}_m)\|_2^2}{\sigma_w^2} + C,
\end{equation}
where $C = M N \log(\pi \sigma_w^2)$ is a constant independent of $\bar{\mathbf{h}}$. Substituting this back into the MAP minimization problem, we get
\begin{equation}
\bar{\mathbf{h}}_{\text{MAP}} = \arg\min_{\bar{\mathbf{h}}} \left( \sum_{m=1}^{M} \frac{\|\hat{\mathbf{h}}(\boldsymbol{\zeta}_m) - \bar{\mathbf{h}}(\boldsymbol{\zeta}_m)\|_2^2}{\sigma_w^2} - \log p(\bar{\mathbf{h}}) \right).
\end{equation}
Multiplying the objective function by the positive constant $\frac{\sigma_w^2}{2}$ does not change the minimizer $\bar{\mathbf{h}}$. This transforms the MAP estimation problem into the canonical form of a regularized optimization problem:
\begin{equation}
\begin{split}\label{MAP} 
\bar{\mathbf{h}}_{\text{MAP}} = \arg\min_{\bar{\mathbf{h}}} \left( \sum_{m=1}^{M} \frac{1}{2} \|\hat{\mathbf{h}}(\boldsymbol{\zeta}_m) - \bar{\mathbf{h}}(\boldsymbol{\zeta}_m)\|_2^2 - \frac{\sigma_w^2}{2} \log p(\bar{\mathbf{h}}) \right),
\end{split}
\end{equation}
where the squared $\ell_2$-norm term acts as a data fidelity term.}

This final expression provides that any channel estimator can be interpreted as a solver for this problem. The first term enforces fidelity to the measurements, while the second term $-2\sigma_w^2 \log p(\bar{\mathbf{h}})$, acts as a regularizer that is fundamentally defined by the choice of the prior $p(\bar{\mathbf{h}})$.
Consequently, the choice of the estimator is fundamentally governed by the choice of the prior $p(\bar{\mathbf{h}})$, which represents our knowledge or assumptions about the channel structure. The model-based methods in Section III make explicit, but often restrictive, prior assumptions, e.g., a strict sparsity prior for OMP estimator. To motivate a more general and physically-grounded prior that can serve as the foundation for a learning-based approach, we now analyze the inherent spectral properties of the FIM channel model itself.
As defined in Eq. \eqref{channel}, the FIM channel is a linear superposition of $S=L \times G$ steering vectors, each corresponding to a distinct path angle $(\theta_s, \phi_s)$. The response of the $n$-th antenna to the $s$-th path can be expressed as a complex exponential of its spatial coordinates $(x_n, z_n)$:
\begin{equation}
[\mathbf{a}(\boldsymbol{\zeta}, \theta_s, \phi_s)]_n = e^{j\left(k_{x,s}x_n + k_{z,s}z_n + \frac{2\pi}{\lambda}\zeta_n \sin\theta_s \sin\phi_s\right)},
\end{equation}
where $k_{x,s}$ and $k_{z,s}$ are the spatial frequencies corresponding to the angle $(\theta_s, \phi_s)$. When we take the 2D spatial Fourier transform $\mathcal{F}$ of the  channel vector $\mathbf{h}(\boldsymbol{\zeta})$, each path contributes a component at its specific spatial frequency and is given by
\begin{equation}
\tilde{\mathbf{h}}(\boldsymbol{\zeta}; \omega_x, \omega_z) = \mathcal{F}({\mathbf{h}}(\boldsymbol{\zeta})) = \sum_{s=1}^{S} \alpha_s'(\boldsymbol{\zeta}) \mathcal{F}\left(e^{j\left(k_{x,s} x_n + k_{z,s} z_n\right)}\right),
\end{equation}
where $\alpha_s'(\boldsymbol{\zeta})$ is the effective complex gain of the $s$-th path, which incorporates the original path gain $\alpha_s$ and the spatially-varying phase shift induced by the FIM deformation. 

Since the Fourier transform of a complex sinusoid is a Dirac delta function, the spatial spectrum $\tilde{\mathbf{h}}(\boldsymbol{\zeta}; \omega_x, \omega_z)$ of the channel is non-zero only at a discrete set of $S$ spatial frequencies.
This derivation proves a fundamental property: for any given deformation $\boldsymbol{\zeta}$, the FIM channel is inherently sparse in the spatial frequency domain, which provides a rigorous physical justification for selecting an superior prior $p(\bar{\mathbf{h}})$. A function that is sparse in the frequency domain is a specific instance of a broader class of functions whose Fourier coefficients decay rapidly. Consequently, a powerful and general prior for this problem should favor functions with this spectral decay characteristic. By operating directly in the Fourier domain and inherently focusing on a limited number of modes \cite{li2023fourier}, the FNO architecture is naturally aligned with the physical properties of the FIM channel.

\textbf{Remark 2:} While both the parametric channel estimation framework in Section III-C and the FNO framework align with the inherent spectral sparsity of channel, they operate under fundamentally different assumptions. The typical sparsity channel estimation approaches rely on a hard sparsity prior, enforced through a predefined, discrete dictionary of angular basis functions. Its performance is therefore sensitive to violations of this rigid model, such as basis mismatch errors and the presence of weak paths in rich scattering environments. In contrast, the FNO learns a continuous operator and imposes a more flexible spectral decay bias, forming a soft prior that prioritizes low-frequency components without rigidly assuming all others are zero. This architectural bias allows the FNO to gracefully handle non-ideal and non-strictly-sparse channels and avoid the basis mismatch problem entirely, motivating its proposal as a more robust and general estimation framework.

\subsection{FIM Channel Estimation as Neural Operator Learning}

A naive application of generic black-box models is ill-suited for this problem, as they lack the necessary inductive biases to handle the continuous deformation space, disregard the underlying wave physics, and suffer from mesh-dependence, making them impractical and data-inefficient \cite{cui2025overview}.
In this work, we reformulate the FIM channel estimation problem as learning an operator $\mathcal{G}$ that maps the channel response function at a given set of pilot deformations $\{\boldsymbol{\zeta}_m\}_{m=1}^{M}$, along with a query for a target deformation, to the channel vector at that target deformation.
The goal is to learn the operator $\mathcal{G}$ such that for any arbitrary and continuous target deformation $\mathbf{\boldsymbol{\zeta}}_{\text{target}}$, we have the corresponding channel estimate  
$\hat{\mathbf{h}}(\mathbf{\boldsymbol{\zeta}}_{\text{target}}) $ as
\begin{equation}
\begin{split}\label{vg_condensed} 
\hat{\mathbf{h}}(\mathbf{\boldsymbol{\zeta}}_{\text{target}}) = \mathcal{G}\left(\{\hat{\mathbf{h}}(\boldsymbol{\zeta}_m)\}_{m=1}^{M}, \mathbf{\boldsymbol{\zeta}}_{\text{target}}\right).
\end{split}
\end{equation}

We develop the FNO to construct an efficient operator $\mathcal{G}$, which can directly processes the continuous function parameterized by $\mathbf{\boldsymbol{\zeta}}_{\text{target}}$ as an input to produce another function, i.e., the channel vector $\hat{\mathbf{h}}(\mathbf{\boldsymbol{\zeta}}_{\text{target}})$ defined over the $N$ antennas. This perfectly matches the fundamental channel estimation challenge for FIM systems, i.e., prediction in a continuous space. Hence, the FNO provides a powerful and efficient parameterization for this operator $\mathcal{G}$. 

\begin{figure}[t]
	\centerline{\includegraphics[width=3in]{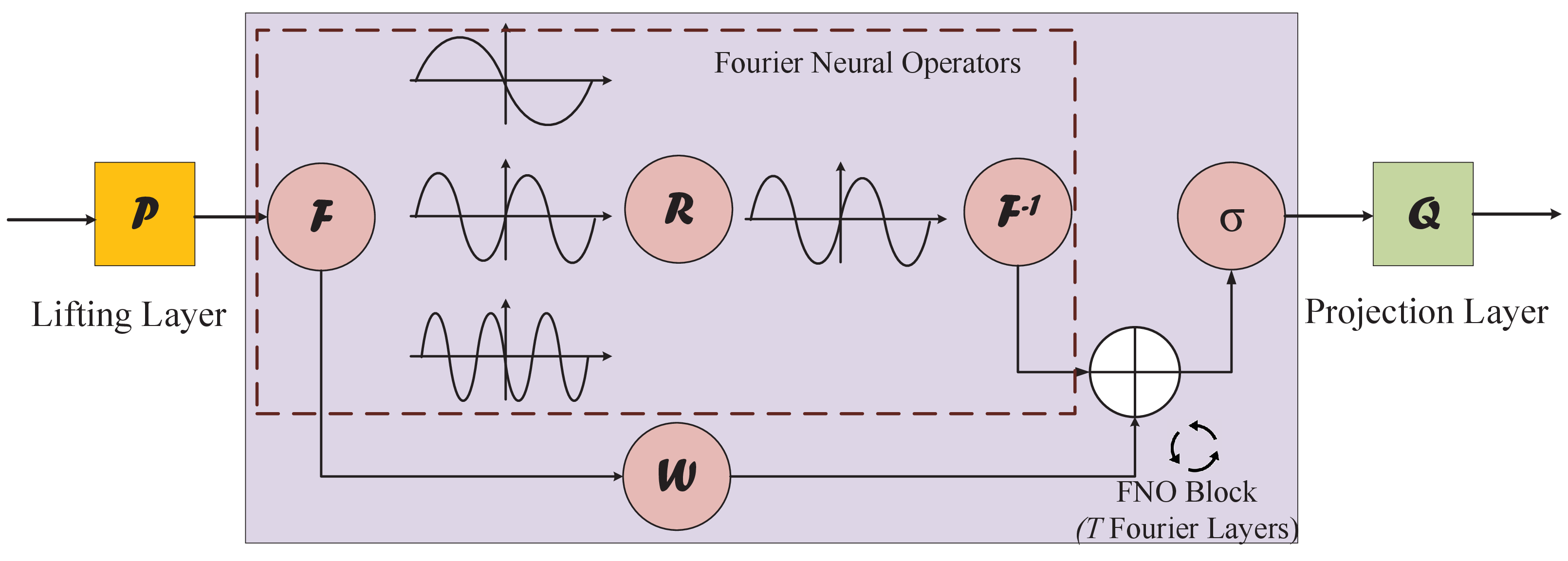}}
	 \caption{Fourier neural operator (FNO) architecture for FIM channel estimation.}
	\label{FNO}
\end{figure}

In Fig.~\ref{FNO}, a typical FNO architecture is presented, which consists of three main components: a lifting layer, a series of Fourier layers for iterative updates in a latent space, and a projection layer. Given that the FIM channel is determined by the superposition and interference of electromagnetic waves, the internal mechanisms of FNO are highly compatible with the problem physics, which efficiently learn systems governed by physical laws.

\subsubsection{Lifting Layer} The input to the FNO is a function defined on the spatial domain of the FIM, which is discretized by $N$ antenna elements. For each antenna element $n$, we construct a high-dimensional feature vector $\mathbf{v}_{\text{in}}(n)\in \mathbb{R}^{3M+1}$ that encapsulates all available information: the measured channel responses and the FIM shape information, ans can be defined as
\begin{equation}
\begin{split}\label{vg_condensed} 
\mathbf{v}_{\text{in}}(n) = &\{ \Re\{\hat{\mathbf{h}}(\mathbf{\boldsymbol{\zeta}}_1)_n\}, \Im\{\hat{\mathbf{h}}(\mathbf{\boldsymbol{\zeta}}_1)_n\}, \dots, \\
&\Re\{\hat{\mathbf{h}}(\mathbf{\boldsymbol{\zeta}}_{M})_n\}, \Im\{\hat{\mathbf{h}}(\mathbf{\boldsymbol{\zeta}}_{M})_n\}, [\mathbf{\boldsymbol{\zeta}}_1]_n, \dots, \\
&[\mathbf{\boldsymbol{\zeta}}_{M}]_n, [\mathbf{\boldsymbol{\zeta}}_{\text{target}}]_n\}^T.
\end{split}
\end{equation}

This input function $\mathbf{v}_{\text{in}}$ is then lifted by a pointwise neural network $\mathcal P$ to a higher-dimensional latent function space, $\mathbf{v}_0(n) = \mathcal P\left(\mathbf{v}_{\text{in}}(n)\right) \in \mathbb{R}^{d_v}$, where $d_v$ is the latent dimension (width) of the FNO, and $\mathcal P$ is implemented as a shallow multi-layer perceptron. 
The lifting layer $\mathcal P$ then maps this input feature function to a higher-dimensional latent function space $\mathbf{v}_0(n) \in \mathbb{R}^{d_v}$. This step enables the model to be conditioned on the specific target deformation $\mathbf{\boldsymbol{\zeta}}_{\text{target}}$ for which the channel is being queried.

\subsubsection{Fourier Layers} The core of the FNO consists of a sequence of $T$ Fourier layers. Each layer $t$ updates the latent representation $\mathbf{v}_t$ to $\mathbf{v}_{t+1}$ according to
\begin{equation}
\begin{split}\label{vg_1} 
\mathbf{v}_{t+1}(n) = \sigma\left( \mathcal{W}_t\mathbf{v}_t(n) + \left(\mathcal{K}(\mathbf{v}_t;\mathbf{\varphi}_t)\right)(n) \right),
\end{split}
\end{equation}
where $\sigma$ is a non-linear activation function. The update of the Fourier layer is composed of two parallel paths: local transformation $\mathcal{W}_t\mathbf{v}_t(n)$ and global kernel integration $\left(\mathcal{K}(\mathbf{v}_t;\mathbf{\varphi}_t)\right)(n)$. The term $\mathcal{W}_t\mathbf{v}_t(n)$ is a pointwise linear transformation applied at each location $n$ with learnable weight $\mathcal{W}_t$, typically implemented as a 1D convolution with a 1$\times$1 kernel. It captures local and channel-wise relationships. The term $\left(\mathcal{K}(\mathbf{v}_t;\mathbf{\varphi}_t)\right)(n)$ is equivalent to a pointwise multiplication in the Fourier domain:
\begin{equation}
\begin{split}\label{vg_2} 
  \left(\mathcal{K}(\mathbf{v}_t;\mathbf{\varphi}_t)\right)(n) = \mathcal{F}^{-1}\left( \mathbf{R}_{\mathbf{\varphi}_t} \cdot \left(\mathcal{F}(\mathbf{v}_t)\right) \right)(n),
\end{split}
\end{equation}
where $\mathcal{F}$ and $\mathcal{F}^{-1}$ denote the fast Fourier transform (FFT) and its inverse, respectively. $\mathbf{R}_{\mathbf{\varphi}_t}$ is a learnable linear transformation with the network weight $\mathbf{\varphi}_t$ that is applied directly to the Fourier modes of $\mathbf{v}_t$. For efficiency and to encourage learning of smooth solutions, this transformation is applied only to the lowest $k_{\text{max}}$ Fourier modes. The higher modes are truncated and set to zero, which can be expressed as
\begin{equation}
\begin{split}\label{vg_3} 
  \left(\mathbf{R}_{\mathbf{\varphi}_t} \cdot \left(\mathcal{F}(\mathbf{v}_t)\right)\right)_\kappa =
  \begin{cases}
  \mathbf{R}_{\mathbf{\varphi}_t,\kappa} \left(\mathcal{F}(\mathbf{v}_t)\right)_\kappa, & \text{for } |\kappa| \le \kappa_{\text{max}}, \\
  0, & \text{for } |\kappa| > \kappa_{\text{max}},
  \end{cases}
\end{split}
\end{equation}
where the choice of the hyperparameter $\kappa_\text{max}$, referred to as FNO modes in the implementation, is constrained by the number of antennas $N$. This constraint arises from the use of the real FFT on the real-valued latent function $\mathbf{v}_t(n)$. Due to the Hermitian symmetry of the Fourier transform of real signals, an input sequence of length $N$ yields only $\lfloor N/2 \rfloor + 1$ unique, non-redundant frequency modes. Consequently, to prevent an index out-of-bounds error during the selection of the lowest modes, the hyperparameter must satisfy the condition $\kappa_\text{max} \le \lfloor N/2 \rfloor + 1$. This ensures the validity of the operation and highlights a practical consideration when applying the FNO model to FIMs of varying sizes.

\textbf{Remark 3:} 
The Fourier-domain operation allows the FNO to efficiently learn global dependencies across the entire FIM array, granting it the powerful property of mesh-independence. {This distinguishes the FNO from standard convolutional networks, which learn spatially localized kernels that are inherently tied to the resolution of the training grid. The learnable weights of FNO are parameterized in the frequency domain, operating on a fixed number of low-frequency Fourier modes. Since these modes represent continuous basis functions, their meaning is independent of the underlying spatial discretization. By learning in this resolution-agnostic domain, the FNO approximates a continuous physical operator rather than a discrete pattern. Consequently, a model trained on an FIM with $N$ antennas can be directly evaluated on a larger-scale FIM with $N' > N$ antennas.}
This offers immense practical value for deploying solutions across FIM hardware of varying scales and configurations.
{While both the proposed FNO and sparsity-based parametric channel estimation methods operate in a transformed domain related to spatial frequencies, their philosophies and implementations are fundamentally different. The sparsity-based methods solve a sparse linear inverse problem on a discrete grid, while FNO learns a continuous and non-linear operator that is biased towards smooth functions.}


\subsubsection{Projection Layer} After $T$ iterations, the final latent representation $\mathbf{v}_T(n)$ is projected back to the desired output dimension by another pointwise neural network $\mathcal Q$. The output represents the real and imaginary parts of the estimated channel at each antenna, which can be expressed as
\begin{equation}
\begin{split}\label{vg_condensed} 
\left[ \Re\left\{\hat{\mathbf{h}}(\mathbf{\boldsymbol{\zeta}}_{\text{target}})_n\right\}, \Im\left\{\hat{\mathbf{h}}(\mathbf{\boldsymbol{\zeta}}_{\text{target}})_n\right\} \right]^T = \mathcal Q\left(\mathbf{v}_T(n)\right).
\end{split}
\end{equation}
Then, the two output channels are combined to form the final complex channel estimate $\hat{\mathbf{h}}(\mathbf{\boldsymbol{\zeta}}_{\text{target}})$.

\subsection{Hierarchical FNO for Multi-Scale Feature Learning}
\begin{figure}[t]
	\centerline{\includegraphics[width=3in]{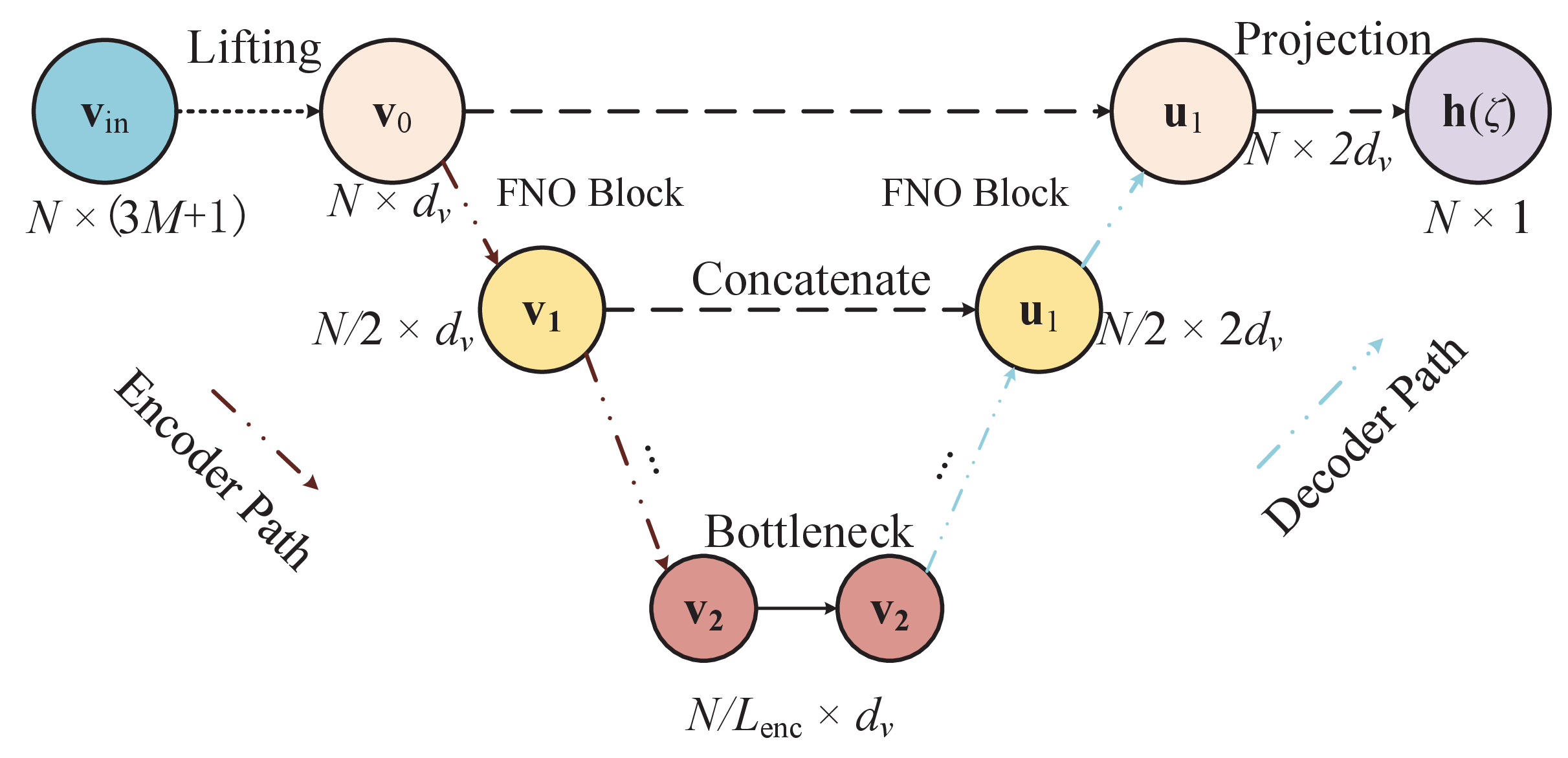}}
	 \caption{Hierarchical Fourier neural operator (H-FNO) architecture for multi-scale FIM channel estimation.}
	\label{HFNO}
\end{figure}
The standard FNO architecture in Section IV-B, while effective, processes information at a single spatial resolution in $T$ Fourier layers. However, the physical deformation of an FIM, and its resulting impact on the channel, often exhibits multi-scale characteristics. These can range from large-scale, low-frequency bending of the entire surface to small-scale, high-frequency ripples on a local level. A single-scale model may struggle to efficiently capture this full spectrum of features. To address this, we propose an effective H-FNO architecture that integrates the multi-scale analysis capabilities of the U-shaped network architecture with the powerful Fourier-domain learning of FNO \cite{ronneberger2015u}. The H-FNO is designed to learn channel features across a hierarchy of spatial resolutions simultaneously, enabling a more comprehensive understanding of the shape-to-channel mapping.

As illustrated in Fig.~\ref{HFNO}, the H-FNO architecture consists of an encoder path, a bottleneck, and a decoder path, connected by skip connections.

\subsubsection{Encoder Path} The encoder progressively extracts more abstract, large-scale features by reducing the spatial resolution. Starting with the lifted feature function $\mathbf{v}_0(n) \in \mathbb{R}^{d_v}$ defined on the $N$ antenna elements, the $l$-th level of the encoder first applies an FNO block and then performs a pooling operation. An FNO block $\mathcal{B}_{\text{FNO}}$ consists of a Fourier layer followed by a non-linear activation $\sigma$, which is given by
\begin{equation}
\mathbf{v}'_l = \mathcal{B}_{\text{FNO}}(\mathbf{v}_{l-1}) = \sigma\left( \mathcal{W}_l\mathbf{v}_{l-1} + \mathcal{K}_l(\mathbf{v}_{l-1};\mathbf{\phi}_l) \right).
\end{equation}
The output is then downsampled by an average pooling operator $\mathcal{A}$, i.e., $\mathbf{v}_l = \mathcal{A}(\mathbf{v}'_l)$. {This operation halves the spatial resolution of the feature map while keeping its channel dimension $d_v$ constant.}
This process is repeated for $L_{\text{enc}}$ levels, resulting in a sequence of latent representations $\{\mathbf{v}'_1, \mathbf{v}'_2, \dots, \mathbf{v}'_{L_{\text{enc}}}\}$ with progressively decreasing spatial dimensions ($N, N/2, \dots, N/2^{L_{\text{enc}}-1}$).

\subsubsection{Bottleneck} At the lowest spatial resolution, a final FNO block is applied to perform deep feature extraction on the most abstract representation $\mathbf{v}_{\text{bottle}} = \mathcal{B}_{\text{FNO}}(\mathbf{v}_{L_{\text{enc}}})$.

\subsubsection{Decoder Path} The decoder path reconstructs the high-resolution channel estimate by progressively upsampling the feature maps and integrating finer details from the encoder path via skip connections. The $l$-th level of the decoder first upsamples the representation from the previous level $\mathbf{u}_{l-1}$, using an operator $\mathcal{U}$ based on linear interpolation:
\begin{equation}
\mathbf{u}'_l = \mathcal{U}(\mathbf{u}_{l-1}),
\end{equation}
This upsampled feature map is then concatenated with the corresponding feature map from the encoder path, $\mathbf{v}'_{L_{\text{enc}}-l+1}$, through a skip connection. This concatenation is crucial as it re-introduces high-resolution spatial details that may have been lost during downsampling, which is given by
\begin{equation}
\mathbf{u}_{l, \text{cat}} = \left[ \mathbf{u}'_l ; \mathbf{v}'_{L_{\text{enc}}-l+1} \right],
\end{equation}
where $\left[ \cdot ; \cdot \right]$ denotes concatenation along the channel dimension. {Specifically, the upsampled feature map with $d_v$ channels is combined with the feature map from the encoder path, resulting in a fused feature map with a doubled channel dimension of $2d_v$}. Finally, another FNO block is applied to fuse these multi-scale features $\mathbf{u}_l = \mathcal{B}_{\text{FNO}}(\mathbf{u}_{l, \text{cat}})$. After $L_{\text{enc}}$ decoding steps, the final high-resolution latent representation is passed through the projection layer $\mathcal Q$ to produce the channel estimate.

By explicitly processing information at multiple scales and fusing features via skip connections, the H-FNO architecture is better equipped to model the complex, multi-faceted relationship between the physical deformation of FIM and the wireless channel.
Let $\Theta$ be the set of all learnable parameters in the H-FNO. The training objective of H-FNO is to minimize the mean squared error (MSE) loss, averaged over the data distribution, which can be expressed as
\begin{equation}
\begin{split}\label{Loss} 
\mathcal{L}(\Theta) = \frac{1}{B_s} \sum_{i=1}^{B_s} \left\| \mathcal{G}_{\Theta}\left(\{\hat{\mathbf{h}}^{(i)}(\boldsymbol{\zeta}_m)\}_{m=1}^{M}, \mathbf{\boldsymbol{\zeta}}_{\text{target}}^{(i)}\right) - \mathbf{h}^{(i)}(\mathbf{\boldsymbol{\zeta}}_{\text{target}}^{(i)}) \right\|_2^2,
\end{split}
\end{equation}
where $B_s$ denotes the input batchsize in the training stage.

As illustrated in Section VI-A, the proposed H-FNO $\mathcal{G}_{\Theta}$ can be interpreted as a learned approximation to the MAP estimator for the FIM channel estimation problem, where the underlying prior imposed by the H-FNO architecture aligns with the inherent spectral properties of the physical channel model. {The mechanism of truncating high-frequency Fourier modes in Eqs. \eqref{vg_1}-\eqref{vg_3} imposes a strong spectral bias on the solution. This architectural choice acts as a powerful structural regularizer, forcing the learned channel function to be inherently smooth and band-limited. This approach is conceptually analogous to the role of the prior in a MAP framework, as both methods effectively constrain the solution to a physically plausible function space.}
{Specifically, the GP is a collection of random variables, any finite number of which have a joint Gaussian distribution \cite{williams2006gaussian}. In this paper, we regard the channel function ${h}(\boldsymbol{\zeta})$ as a GP, mapping deformation $\boldsymbol{\zeta}$ to channel response. The FNO truncation of high-frequency Fourier modes implements a smoothness constraint, effectively acting as a hard spectral prior, whereas the GP uses a soft probabilistic prior via the kernel. Both approaches favor band-limited channel functions, but the FNO achieves this through architectural design rather than probabilistic modeling.}

For the data fidelity term of MAP, i.e., the first term in Eq. \eqref{MAP}, the H-FNO is trained end-to-end by minimizing the MSE loss in Eq. \eqref{Loss}, which is a direct data-driven approximation of the data fidelity term. For the second regularization term in Eq. \eqref{MAP} that penalizes high-frequency channel content,
the H-FNO implements this regularization structurally. {Its architectural truncation of Fourier modes beyond $\kappa_{\text{max}}$ is not merely a soft penalty but a hard constraint. This is structurally equivalent to imposing a MAP prior $p(\bar{\mathbf{h}})$, where the probability of any function containing these higher frequencies is explicitly zero. Consequently, the network is architecturally incapable of representing non-band-limited functions, forcing the training process to find a solution entirely within a physically-motivated function space that adheres to this strong spectral bias.}
Within the preserved modes, the learnable transformation $\mathbf{R}_{\varphi}$ can be interpreted as learning the inherent structure of the channel function space.

\textbf{Remark 4:} In the popular deep learning-enabled channel estimators, the deep neural network does not require an explicit prior, which implicitly imposes a learned structural prior on the function space of possible channels as a generic function approximator. However, the proposed H-FNO is not a closed box but a principled and data-driven method for finding a powerful approximation $\mathcal{G}_{\Theta}$, to the true MAP estimator. {The FNO architecture is explicitly designed to search within a function space constrained by a physically-motivated spectral bias, which is structurally enforced by truncating high-frequency components in the Fourier domain.} This theoretical synergy between the inductive biases of the proposed H-FNO and the physical characteristics of FIM provides a rigorous justification for its application.

{\subsection{Complexity Analysis}
\subsubsection{Computational Complexity}In the proposed H-FNO architecture, the primary computational bottleneck is the FFT operation within each Fourier layer. For an FIM with $N$ radiating elements and a latent channel dimension of $d_v$, a single FFT has a complexity of $\mathcal{O}(N \log N)$. While the H-FNO processes features across multiple spatial resolutions, the overall cost is dominated by the layers with the highest resolution. Therefore, the total inference complexity is approximately $\mathcal{O}(L d_v  N \log N)$,
where $L=2 {L_{\text{enc}}} + 1$ is the total number of Fourier layers in the network. This demonstrates that the complexity scales quasi-linearly with the FIM size, making the H-FNO efficient even for large arrays.} {The training complexity of the proposed H-FNO architecture for a single epoch depends on the cost of the forward and backward passes, repeated for all batches. As the complexity of backpropagation is a small constant multiple of the forward pass, the total training complexity per epoch is given by $\mathcal{O}\left(\frac{N_s}{B_s} L d_v N \log N\right)$, where $N_s$ is the total number of training samples and $B_s$ is the batch size in the training stage.}
{\subsubsection{Sample Complexity}For the considered channel estimation problem in FIM systems, the sample complexity, i.e., the number of pilot measurements $M$ required for generalization, is determined by the intrinsic complexity of the channel function $\mathbf{h}(\boldsymbol{\zeta})$. This complexity relates to the function smoothness in the deformation space. Following principles analogous to the Nyquist-Shannon theorem, a smoother function requires fewer measurements to be accurately characterized. This concept is directly embodied in the H-FNO design. The number of retained Fourier modes $\kappa_{\text{max}}$, acts as an architectural prior on the smoothness of the learned operator. A smaller $\kappa_{\text{max}}$ constrains the model to represent smoother functions, thereby aligning the model's capacity with the information available from a limited number of pilot measurements $M$.}

\section{Numerical Results}
In this section, we provide extensive numerical results to validate the effectiveness of the proposed H-FNO framework by comparing it against several model-based benchmarks, and also present the interpretability analysis of H-FNO.
\subsection{Parameter Setups and Dataset Construction}
In the simulations, we consider the BS is located at a fixed position of (0, 0, 10) m, while the $K$ users are randomly and uniformly distributed within a rectangular volume defined by the diagonal corner coordinates of (-10, 10, 0) m and (10, 30, 0) m. Unless otherwise specified, we set the FIM size to $N = N_x \times N_z = 8 \times 8$, the number of pilot shapes to $M=16$, the number of users to $K=4$, the maximum deformation range to $\bar{\zeta} = \lambda/2$, the number of scattering clusters to $L=5$ with $G=6$ paths per cluster, the path loss exponent to $v=2.2$, and the carrier frequency to $f_c = 28$ GHz.
The performance of all algorithms is evaluated using the normalized mean squared error (NMSE), defined as $\text{NMSE}=\mathbb{E}\{ {{||\hat{\mathbf{h}}_k(\boldsymbol{\zeta_{\text{target}}})-\mathbf{h}_k(\boldsymbol{\zeta_{\text{target}}})||}_{F}^{2}/{||\mathbf{h}_k(\boldsymbol{\zeta}_{\text{target}})||_{F}^{2}}}\}$. 

In the training stage of the proposed H-FNO, a dataset of $N_s = 20,000$ samples is synthetically generated, with $N_v = 1000$ of these samples being reserved for the validation set. For each sample, a unique channel environment is realized based on the clustered multipath model, and then $M$ noisy pilot observations are generated, where the corresponding noise-free channel vector $\mathbf{h}_k(\boldsymbol{\zeta}_{\text{target}})$ serves as the ground-truth label. To enhance the model robustness, the SNR for each training batch is randomly selected from \{0, \ldots, 20\} dB. The model is trained for $E_s=50$ epochs with a batch size of $B_s = 64$. For each SNR point under performance testing, the NMSE results are averaged over $U=500$ Monte Carlo simulations. In each simulation trial, both a new channel realization and a new random target deformation shape are generated to ensure a robust and fair evaluation. 
For the benchmark algorithms, the number of nearest neighbors for KNN is set to $\bar{k}=5$, and the OMP dictionary is constructed from a discrete grid of $D=256$ angles. For the SBL estimator, the iterative learning process is controlled by a maximum of $100$ iterations and a convergence tolerance of $10^{-4}$. For the H-FNO, we set the latent dimension to $d_v = 64$ and the number of encoder levels to $L_\text{enc} = 2$, with the number of Fourier modes per level set as $\kappa_{l,\text{max}}=16/2^l$, which represent a standard architecture designed to balance model expressiveness with computational efficiency. 

\subsection{Performance Comparison between Different Algorithms}
\begin{figure}[t]
	\centerline{\includegraphics[width=3in]{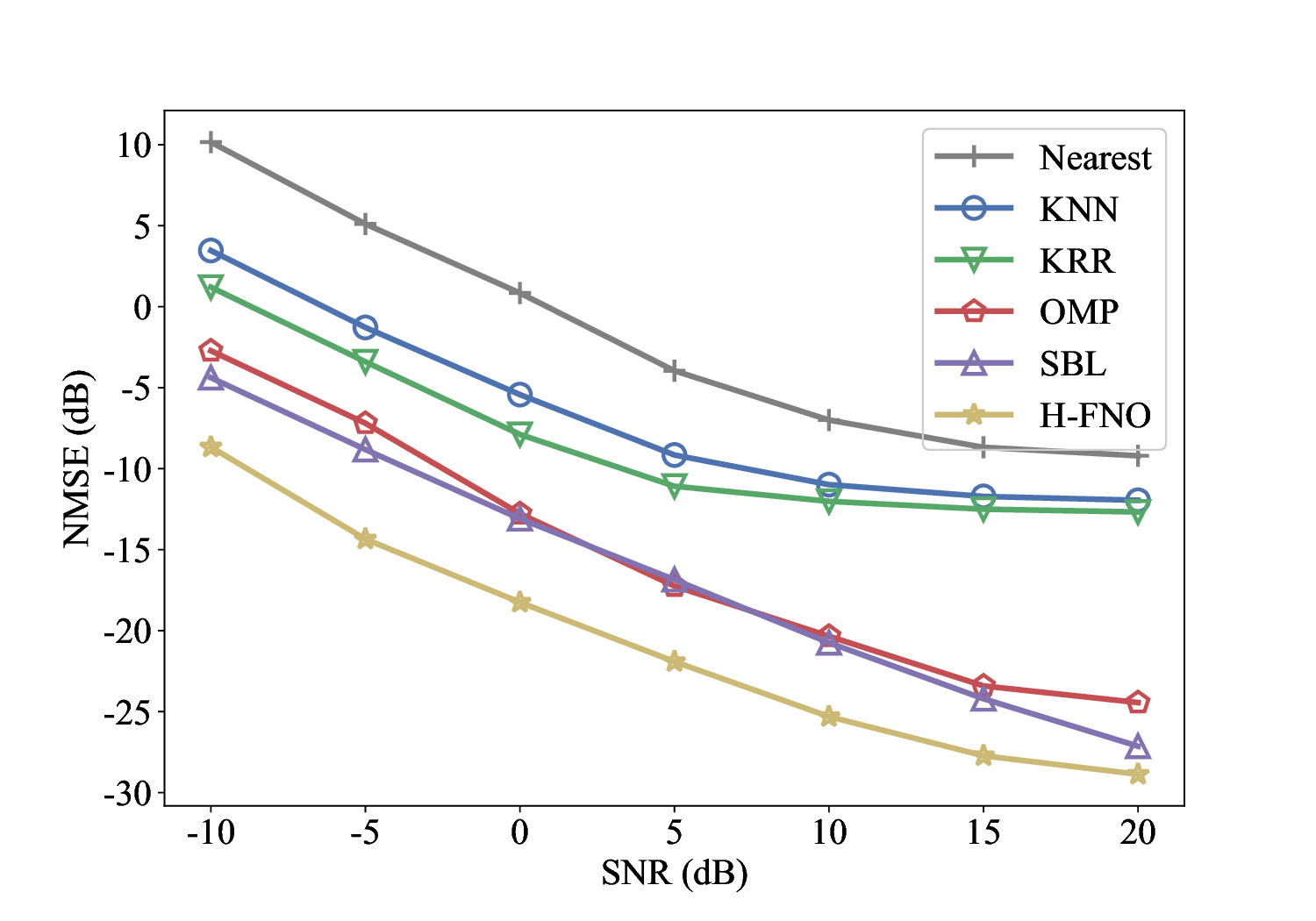}}
	 \caption{NMSE performance against SNR for different channel estimation algorithms.}
	\label{SNR}
\end{figure}
Fig.~\ref{SNR} presents the NMSE performance of different channel estimation algorithms versus the signal-to-noise ratio (SNR), where the channel estimation accuracy of all methods improves with increasing SNR. {Among the conventional model-based methods, the parametric channel estimation schemes, i.e., OMP and SBL, show superior channel estimation accuracy, benefiting from the inherent sparsity of the mmWave channel, while interpolation-based methods provide a stable but less accurate estimation. In particular, the SBL algorithm can offer superior performance compared to OMP by avoiding local minima and providing a probabilistic estimate of the channel parameters.} The proposed H-FNO model consistently and significantly outperforms all other conventional model-based algorithms across the entire SNR range, demonstrating its powerful capability in learning the complex mapping from FIM shapes to channel responses and its robustness against noise. {Note that while the numerical results show the superior accuracy of a learning-based approach, the conventional model-based frameworks retain significant practical value. They are highly interpretable, provide theoretical performance guarantees, and can be deployed without large training datasets or complex offline training pipelines.} 

\begin{figure}[t]
	\centerline{\includegraphics[width=3in]{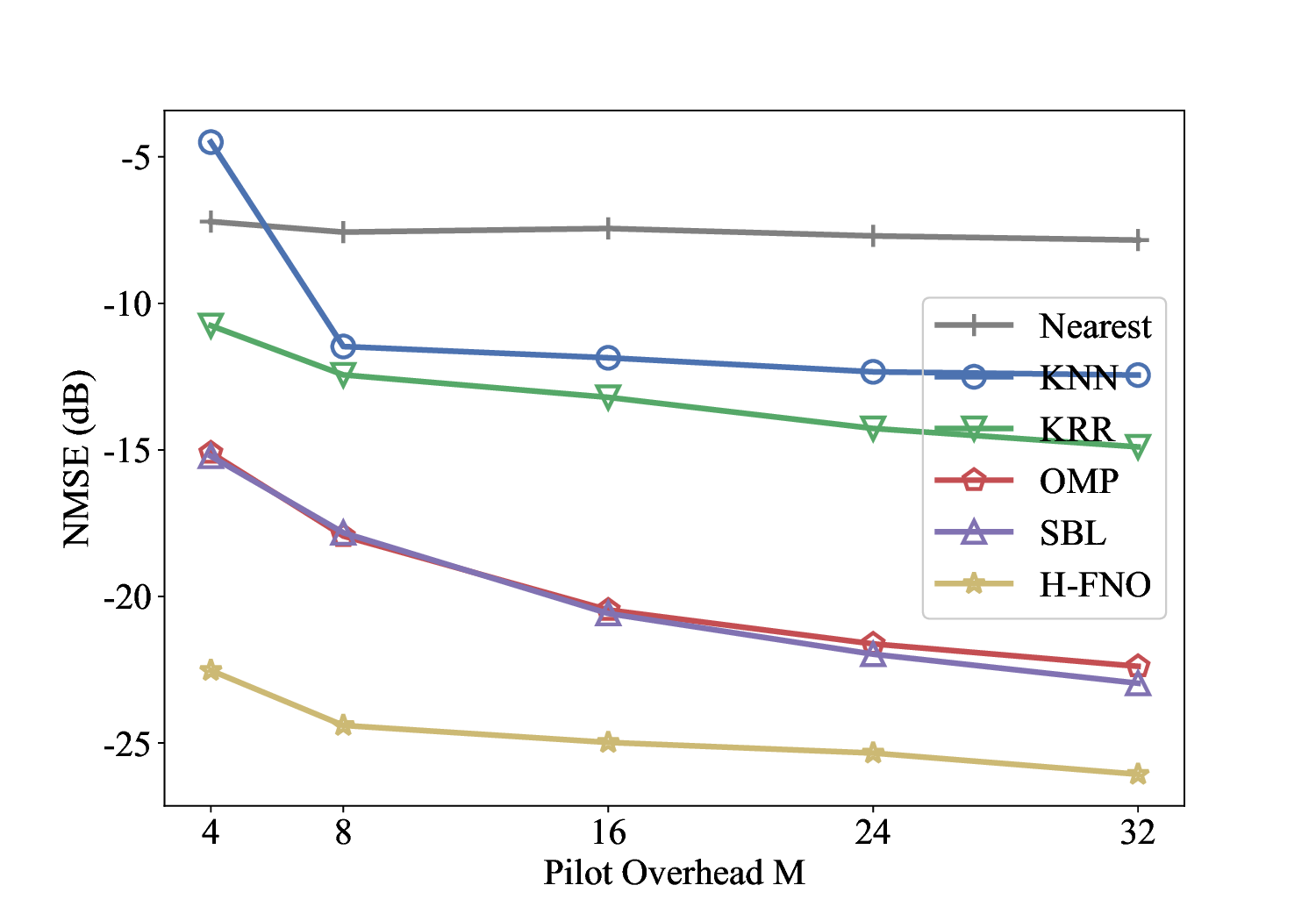}}
	  \caption{NMSE performance against pilot overhead $M$ for different channel estimation algorithms.}
	\label{Pilots1}
\end{figure}
Fig.~\ref{Pilots1} illustrates the NMSE performance against the pilot overhead $M$ at SNR=10 dB, i.e., the number of FIM deformation shapes used for channel probing. This results highlights the critical trade-off between estimation accuracy and pilot overhead. The proposed H-FNO model demonstrates remarkable data efficiency, achieving a very low NMSE with only a small number of pilot measurements. While the performance of all algorithms improves with more pilots, conventional methods require a significantly larger pilot overhead $M$ to reach a comparable accuracy level. This showcases the ability of H-FNO to learn a generalizable channel operator, enabling low-overhead channel estimation.

\begin{figure}[t]
	\centerline{\includegraphics[width=3in]{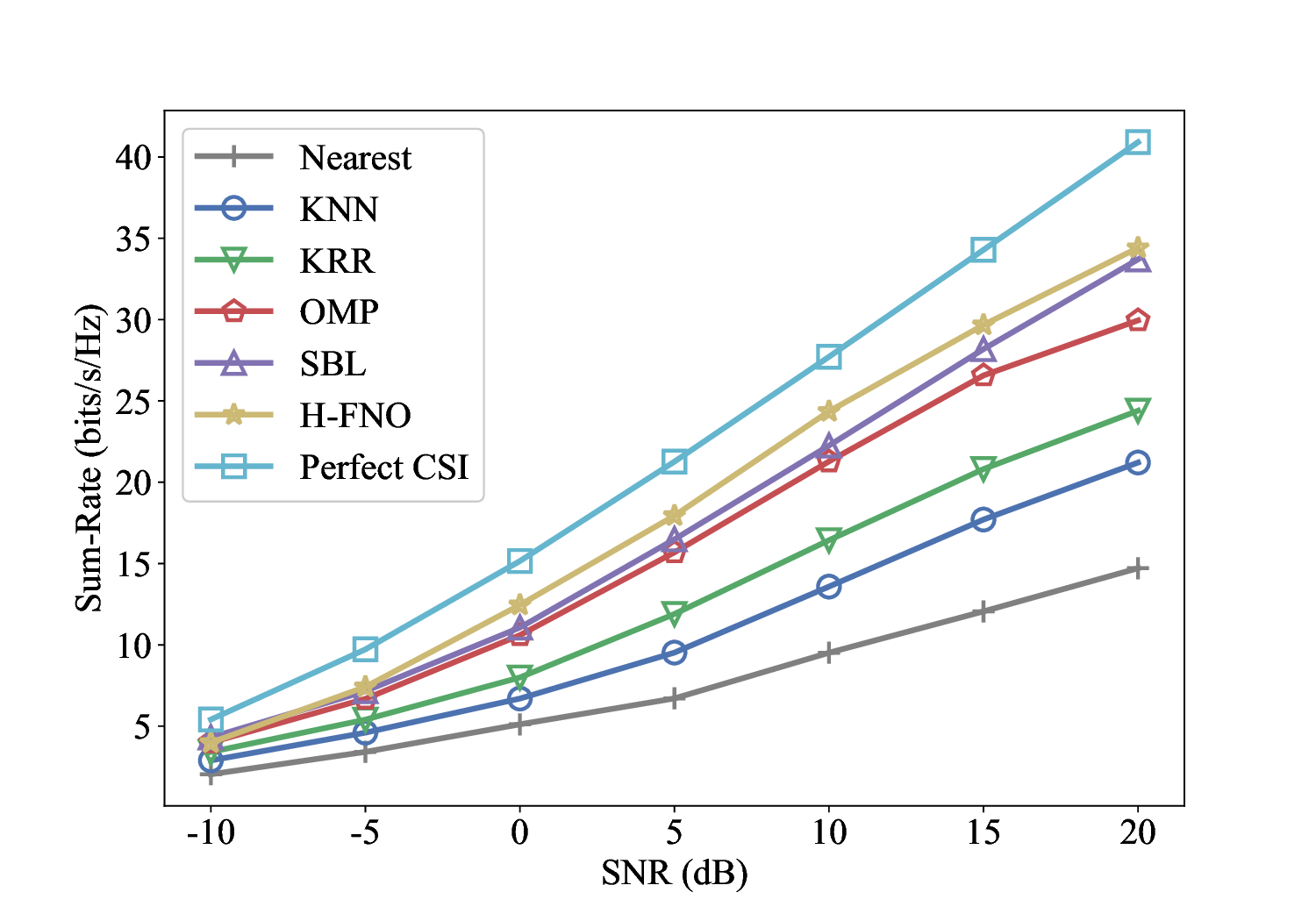}}
	 \caption{Sum-rate performance against SNR for different channel estimation algorithms.}
	\label{Rate}
\end{figure}

{Fig.~\ref{Rate} illustrates the sum-rate performance achieved when using the estimated channels from different algorithms for zero-forcing beamforming \cite{10850658}. We can observe that the proposed H-FNO framework consistently outperforms all model-based benchmarks, achieving a sum-rate significantly closer to the theoretical upper bound provided by perfect CSI. This demonstrates that the superior channel estimation accuracy of H-FNO, as demonstrated by its lower NMSE, directly translates into substantial gains in the overall spectral efficiency of FIM assisted multi-user systems.}

\subsection{Generalization Performance of the Proposed H-FNO}
\begin{figure}[t]
	\centerline{\includegraphics[width=3in]{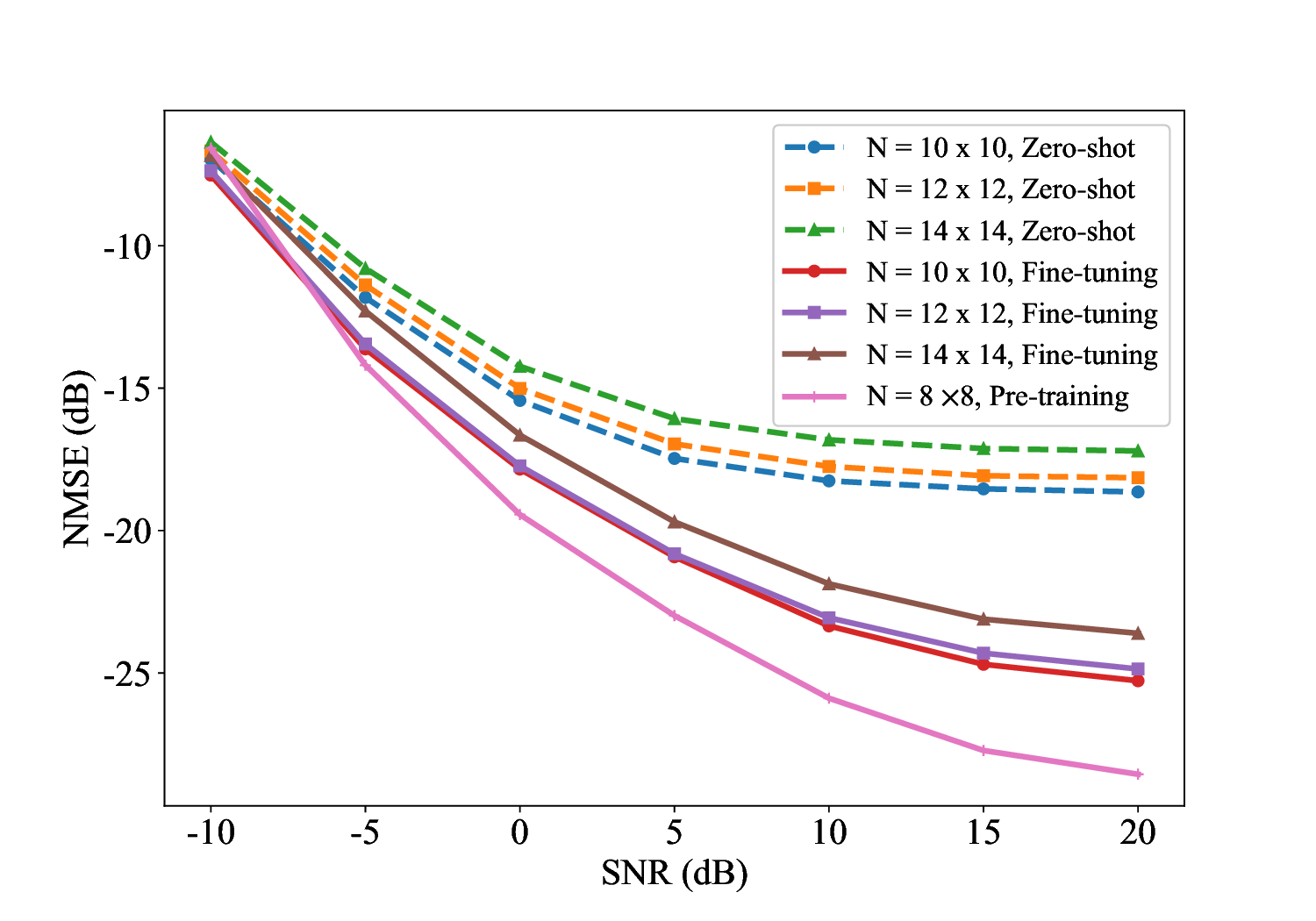}}
	 \caption{NMSE performance of different number of antennas $N$ for the proposed H-FNO.}
	\label{Antennas}
\end{figure}
Fig.~\ref{Antennas} evaluates the mesh-independence property and zero-shot learning ability of the proposed H-FNO model by testing its performance across different numbers of FIM antenna elements $N$. First, the results show that a single H-FNO model, trained on an FIM with a specific size $N=8\times8$, can be directly applied to FIMs of various other sizes from $N=10\times10$ to $N=14\times14$ without retraining. The performance remains robust and degrades gracefully, validating the theoretical mesh-independence of the H-FNO architecture. This is a crucial feature for practical deployment, as it allows a single trained model to be compatible with varying FIM configurations. Second, we provided the fine-tuning model for the cases of $N=10\times10$ to $N=14\times14$, where $N_f = 500$ training samples under new FIM configurations are generated to finetune the pretrained model with $E_f = 10$ epochs.  We observe that the proposed HFO model pre-trained on a small-scale FIM array can be rapidly adapted to larger arrays, achieving superior estimation accuracy with only a brief fine-tuning period on a minimal amount of new data. This confirms that the learned physical knowledge is generalizable and can be effectively transferred across different array scales, significantly reducing the data and training requirements for new FIM deployments.


\begin{figure}[t]
	\centerline{\includegraphics[width=3in]{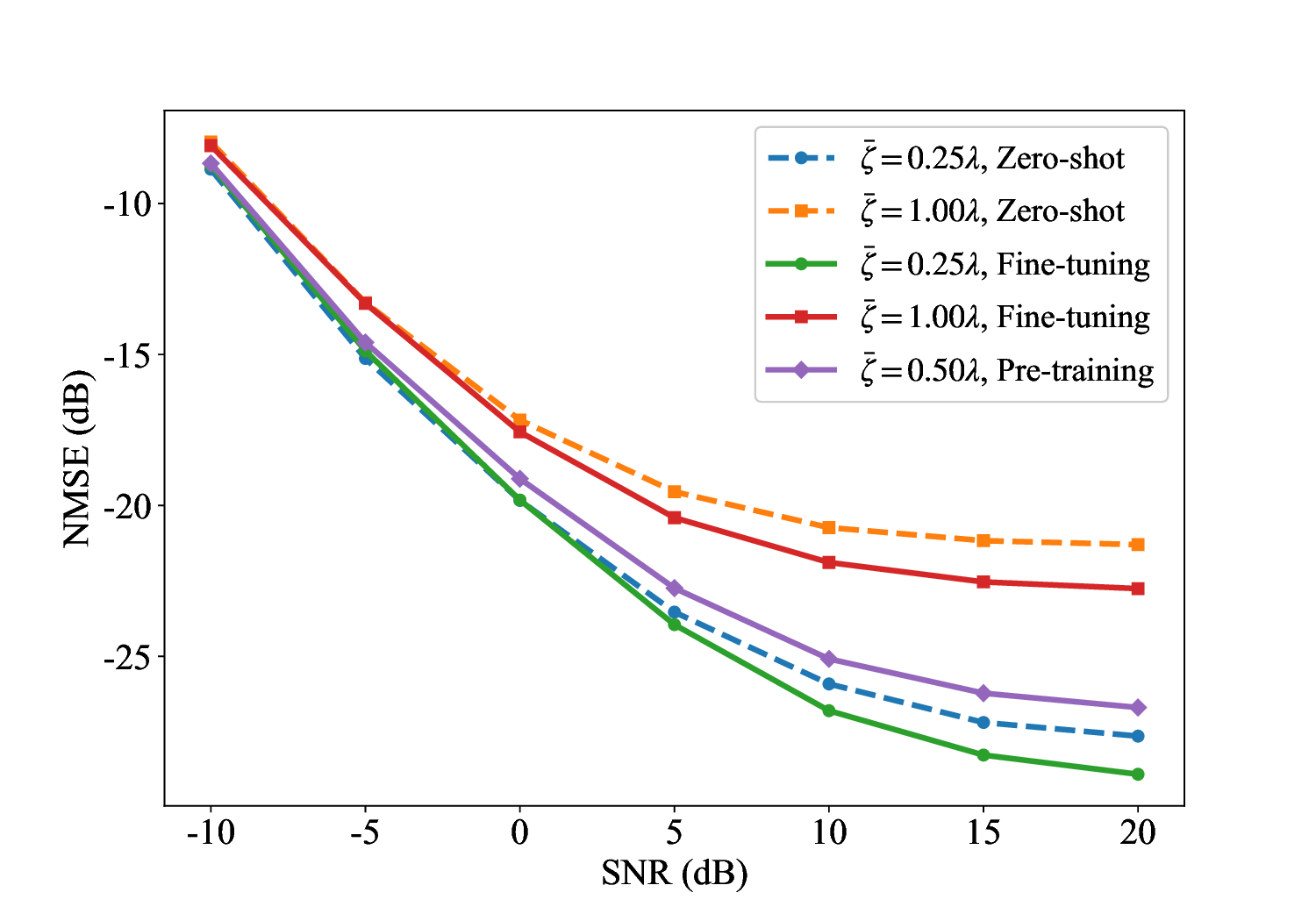}}
	 \caption{NMSE performance of different deformation distances $\bar{\zeta}$ for the proposed H-FNO.}
	\label{Deformation}
\end{figure}

Fig.~\ref{Deformation} validates the zero-shot generalization capability of the proposed H-FNO model across different FIM hardware configurations, specifically varying maximum deformation distances $\bar{\zeta}$. We evaluate a single H-FNO model, pre-trained on a medium deformation distance of $\bar{\zeta} = 0.50\lambda$, on two distinct and challenging scenarios. The smaller range $\bar{\zeta} = 0.25\lambda$ tests the model ability to interpolate, assessing its robustness when the physical diversity of the FIM is less than what it was trained on. Conversely, the larger range $\bar{\zeta} = 1.00\lambda$ tests the model ability to extrapolate, evaluating whether the learned physical principles hold for deformations beyond the scope of its training data. The results demonstrate that the model maintains excellent performance in both the interpolation and extrapolation settings. This indicates that the proposed H-FNO is not simply overfitting to a specific configuration but is learning the underlying physical relationship between deformation and the channel response. 


\begin{figure}[t]
	\centerline{\includegraphics[width=3in]{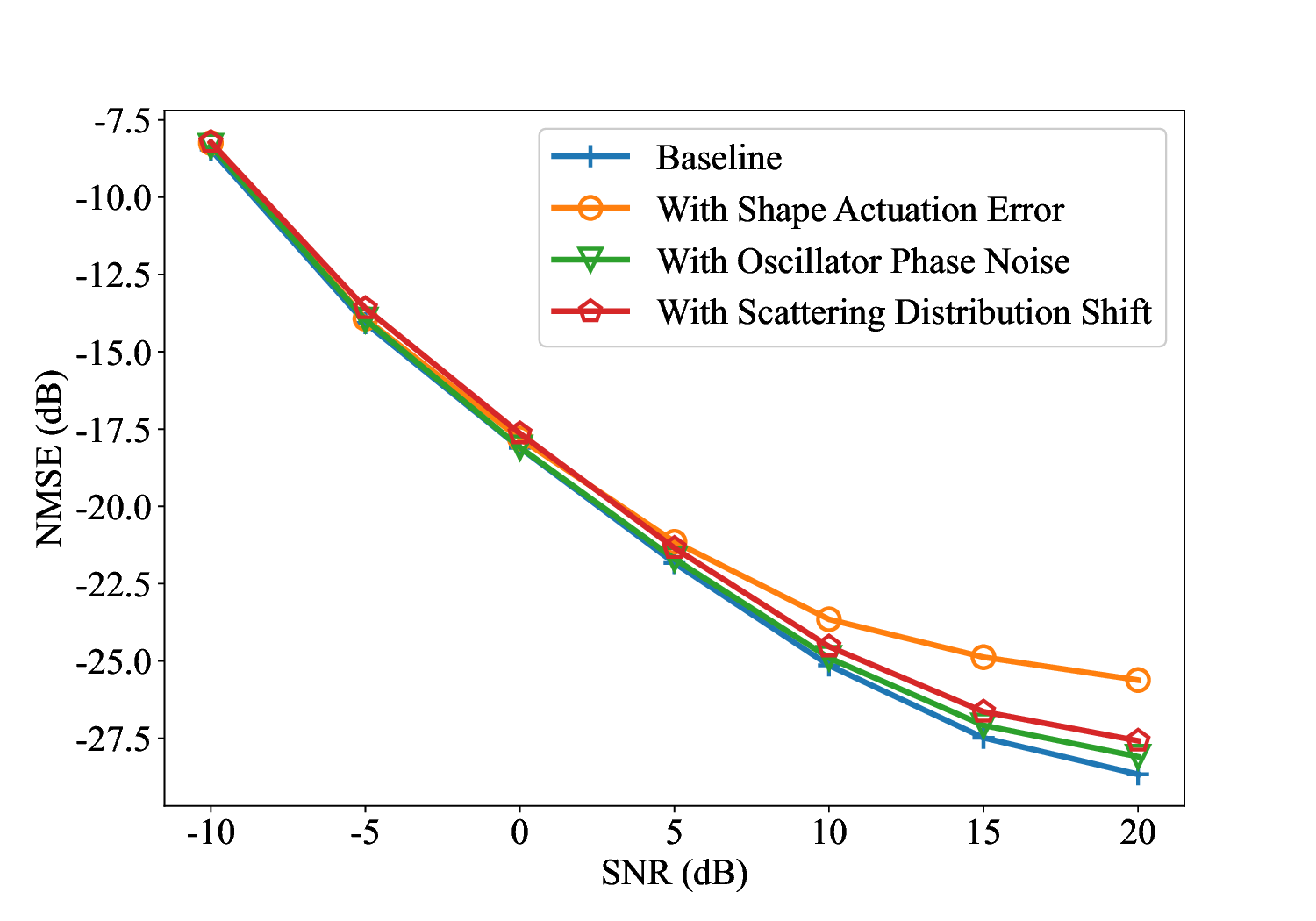}}
	 \caption{Robustness analysis under impairments and distribution shift for the proposed H-FNO.}
	\label{Robust}
\end{figure}

{Fig.~\ref{Robust} evaluates the robustness of the proposed H-FNO framework, where three types of hardware imperfections are introduced during the testing phase: 1) a pilot shape actuation error, modeled by a Gaussian perturbation with a standard deviation set to 5\% of the maximum deformation range $\sigma_{\text{act}} = 0.05 \bar{\zeta}$; 2) oscillator phase noise, introduced as a random phase shift with a standard deviation of 0.05 radians; and 3) a channel distribution shift, where the model, trained on channels with $L=5$ clusters, is tested on a more complex rich-scattering environment with $L=20$ clusters. The results show only a slight performance degradation in the presence of the hardware impairments, demonstrating strong robustness. Furthermore, the model maintains highly effective estimation even under the significant channel distribution shift, which confirms its ability to learn generalizable physical principles rather than merely overfitting to the training statistics.}

\subsection{Interpretability and Visualization of the Proposed H-FNO}

\begin{figure}[t]
	\centerline{\includegraphics[width=3.1in]{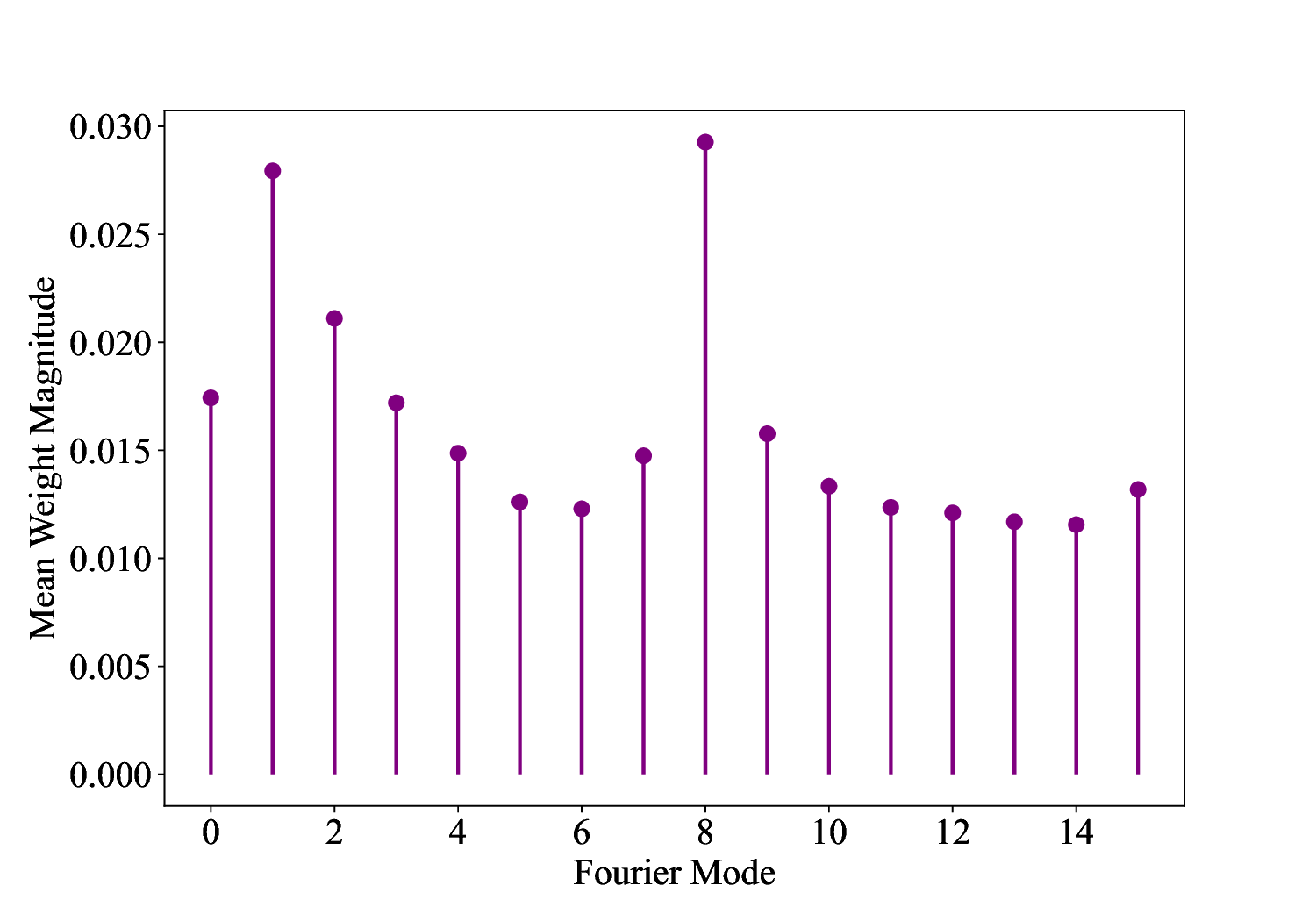}}
	 \caption{Fourier layer weight visualization of the proposed H-FNO.}
	\label{Fourier Weight}
\end{figure}

\begin{figure}[t]
	\centerline{\includegraphics[width=3in]{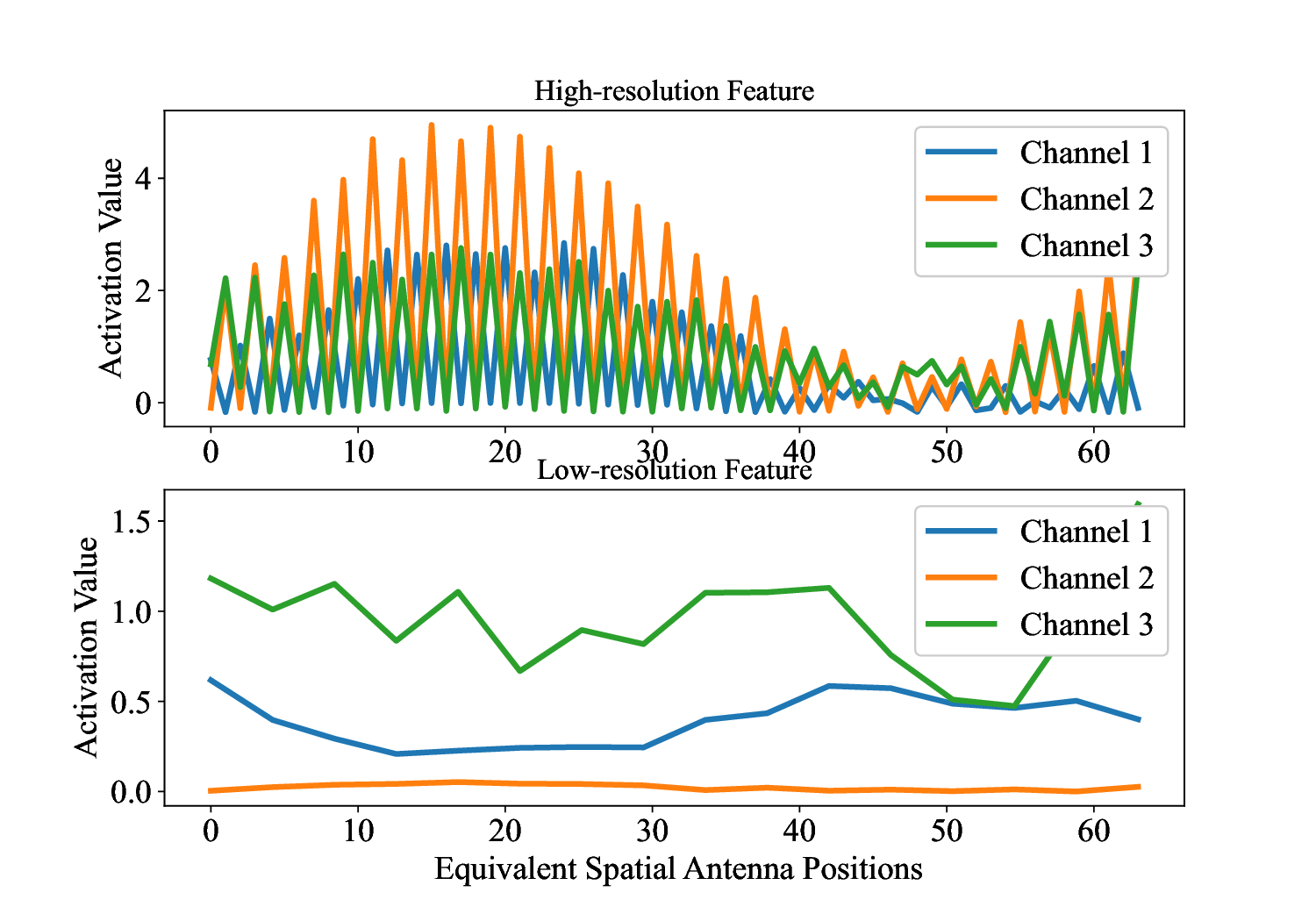}}
	 \caption{Hierarchical feature visualization of the proposed H-FNO.}
	\label{Hierarchical Feature}
\end{figure}

Fig.~\ref{Fourier Weight} visualizes the learned weights of the first Fourier layer within the H-FNO architecture. The result shows the average magnitude of the complex-valued weights $R_{\phi}$, applied to each Fourier mode, which corresponds to a specific spatial frequency. A key observation is that the learned filter is not a simple, isotropic low-pass filter. Instead, it exhibits a more complex, anisotropic characteristic that is perfectly adapted to the 2D physical structure of the FIM array being processed as a 1D vector. The filter assigns high importance to the lowest frequency modes (e.g., modes 0-2), which are essential for capturing the large-scale, slowly varying components of the channel across the entire metasurface. However, the most prominent and insightful feature is the sharp, dominant peak at mode 8. Given that the $8\times8$ FIM array is flattened column-wise, a spatial period of 8 elements corresponds to the fundamental frequency along the second spatial dimension, i.e., the z-axis. The strong peak at mode 8 reveals that the model has autonomously learned that to accurately reconstruct the channel, it must prioritize not only the large-scale trends but also the distinct periodic structure occurring within each column of the FIM.
This demonstrates that the proposed H-FNO learns a physically-grounded and non-trivial set of features from the data representation. Rather than applying a simple smoothing function, it has discovered the anisotropic nature of the problem and developed a specialized filter that is explicitly tailored to the unique geometry of the FIM array.

Fig.~\ref{Hierarchical Feature} provides insight into the multi-scale feature learning enabled by the hierarchical architecture of H-FNO. It compares feature activations from a shallow high-resolution layer (Encoder 1 of H-FNO, $N$=64) with those from the deepest low-resolution layer (Bottleneck of H-FNO, $N/4=16$). The features from the high-resolution encoder are highly oscillatory, capturing the fine-grained details and rapid phase variations of the channel between adjacent antenna elements. In contrast, the features at the bottleneck are significantly smoother, representing the macro-scale trends and the overall envelope of the channel response across the FIM. This result confirms that the H-FNO effectively decomposes the complex estimation problem by learning features at multiple scales


\begin{figure}[t]
	\centerline{\includegraphics[width=3in]{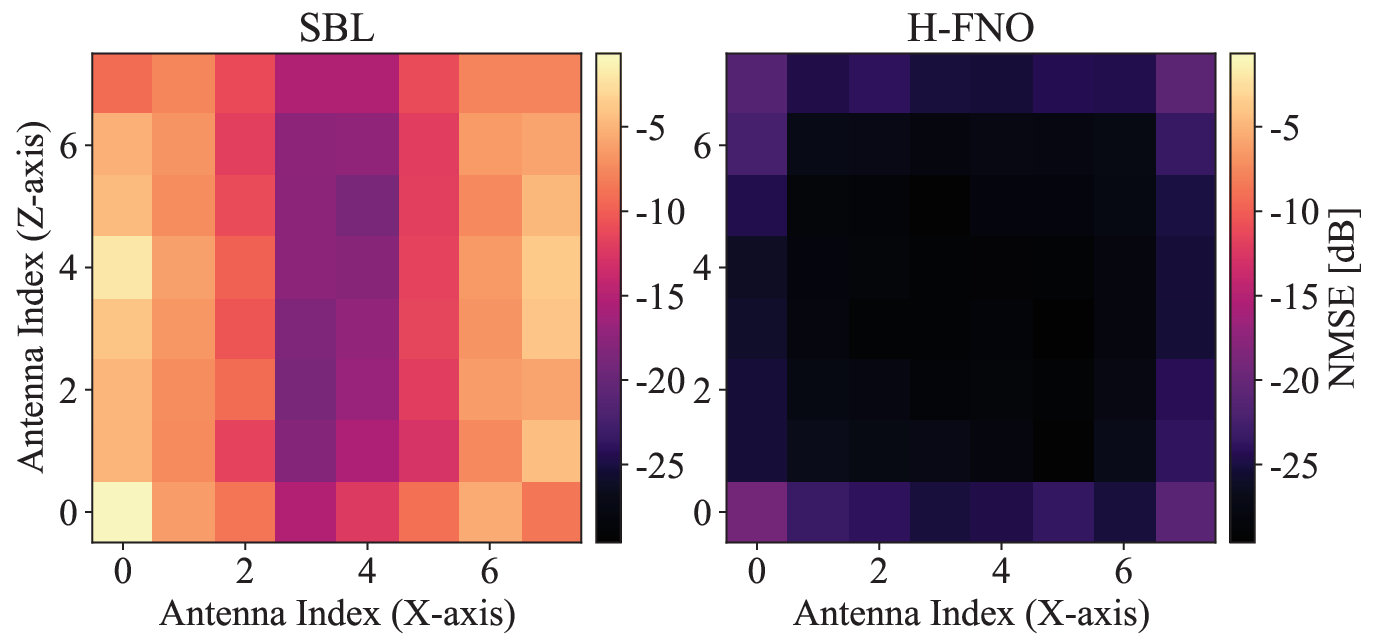}}
	 \caption{Spatial error distribution of per-antenna channel estimation NMSE.}
	\label{Estimation Error}
\end{figure}

\begin{figure}[t]
	\centerline{\includegraphics[width=3.6in]{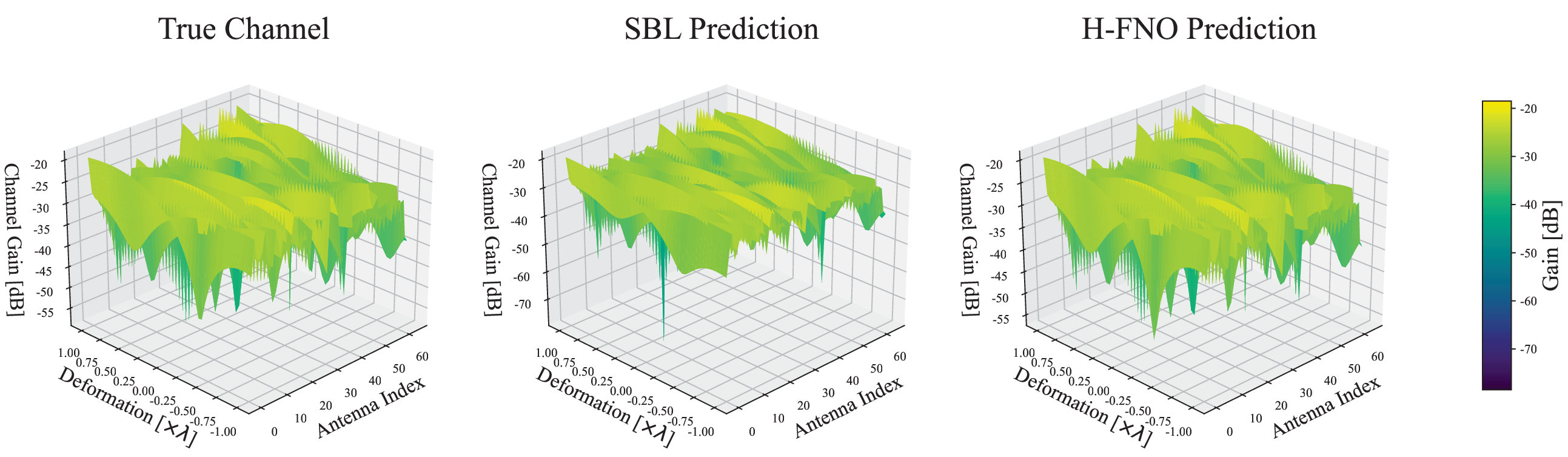}}
	 \caption{3D visualization of channel gain over the joint antenna and deformation domain.}
	\label{2D_Interpolation}
\end{figure}


{Fig.~\ref{Estimation Error} presents a spatial diagnostic of the estimation performance by visualizing the per-antenna NMSE across the $8\times8$ FIM at an SNR of 10 dB. It is observed that while both estimators exhibit higher errors towards the edges of the array, i.e., a common artifact in array processing, the H-FNO demonstrates significantly superior performance. Compared to the SBL algorithm, the proposed H-FNO features a substantially larger dark region, indicating a lower error floor across the central elements. This illustrates that the H-FNO not only achieves a lower overall NMSE but also provides a more uniformly accurate channel estimate across the entire FIM, which is crucial for subsequent signal processing operations.}

Fig.~\ref{2D_Interpolation} presents a 3D visualization of the channel gain landscape as a function of both the antenna index and the continuous deformation distance, {where the channel gain represents the magnitude of the channel response at each individual radiating element, expressed in decibels as $20\log_{10}(|\mathbf{h}_{{{\zeta}_n},n}|)$ (dB).} The leftmost subfigure reveals the complex ground truth surface of the true channel, characterized by numerous peaks and troughs that represent the optimal deformation positions for each antenna. The predicted landscape of H-FNO remarkably reconstructs this intricate surface with high fidelity, correctly identifying the locations and magnitudes of the gain variations. Conversely, the SBL-based prediction landscape is visibly distorted and flattened, failing to capture the deep nulls and high peaks of the true channel. This provides compelling qualitative evidence that the H-FNO has successfully learned a high-fidelity surrogate model for the true physical channel operator, accurately mapping the joint spatial-deformation domain.

\section{Conclusion}
In this paper, we have investigated channel estimation for FIM-aided wireless systems, aiming to address the critical challenge of acquiring the accurate channel across a high-dimensional and continuous deformation space. We have first developed principled model-based frameworks, including interpolation-based, KRR-based and sparsity-based parametric channel recovery, which provide structured solutions to the estimation problem. To push the performance envelope further, we have then proposed a efficient learning-based framework centered on an H-FNO. The proposed H-FNO architecture achieves a superior estimation accuracy and requires significantly lower pilot overhead compared to the conventional model-based techniques. Furthermore, the interpretability analysis have revealed that the proposed H-FNO is not merely a closed-box architecture. Instead, it learns an physically-consistent modeling by capturing multi-scale features and anisotropic spatial structure of the FIM channel. In future works, we will investigate the joint optimization framework of channel estimation and pilot design under the non-ideal deformations and practical aperture losses of FIMs.

\ifCLASSOPTIONcaptionsoff
  \newpage
\fi

\bibliographystyle{IEEEtran}
\bibliography{IEEEabrv,refs_ofdm.bib}
\end{document}